%Paper: gr-qc/9211004
%From: ORTIZ@PIERRE.MIT.EDU
%Date: Wed, 4 Nov 1992 16:00:06 -0500 (EST)
%Date (revised): Wed, 4 Nov 1992 16:27:51 -0500 (EST)
%Date (revised): Wed, 4 Nov 1992 16:33:16 -0500 (EST)
%Date (revised): Wed, 27 Jan 1993 15:45:11 -0500 (EST)

%%%%%%%%%%%%%%%%%%%%%%%%%%%%%%%%%%%%%%%%%%%%%%%%%%%%%%%%%%%%%%%%%
%                                                               %
%   This paper requires the macro JNL.TEX which MUST ITSELF     %
%   INCLUDE the macros REFORDER.TEX and EQNORDER.TEX.           %
%   The version of JNL.TEX available with from hep-th or gr-qc  %
%   with the command "get jnl.tex" includes both these          %
%   macros. If in doubt, use that version of JNL.TEX            %
%                                                               %
%   Figures are available from the authors.                     %
%                                                               %
%%%%%%%%%%%%%%%%%%%%%%%%%%%%%%%%%%%%%%%%%%%%%%%%%%%%%%%%%%%%%%%%%

\input jnl

\tenpoint       % These two lines may be commented out to get
\singlespace    % 12pt double spaced output.

%%%%%%%%%%%% INTERNAL MACROS %%%%%%%%%%%%%%%%%%%%%%%%%%%%%%%%%%%%

\def\d{\delta}
\def\la{\langle}
\def\ra{\rangle}
\def\ria{\rightarrow}
\def\pp{\prime\prime}
\def\x{{\bf x}}
\def\y{{\bf y}}
\def\s{\sigma}
\def\xpp{{\bf x}^{\prime\prime}}
\def\tpp{t^{\prime\prime}}
\def\Xpp{{x^{\mu}}^{\prime\prime}}
\def\X'{{x^{\mu}}^{\prime}}
\def\D{{ \mathrel{\mathop\partial^{\leftrightarrow}} \over \partial X^0} }
\def\G{{\cal G}}
\def\R{{I\kern-0.3em R}}
\def\ula#1{\raise2ex\hbox{$\leftarrow$}\mkern-16.5mu #1}
\def\ura#1{\raise2ex\hbox{$\rightarrow$}\mkern-16.5mu #1}
\def\ulra#1{\raise1.5ex\hbox{$\leftrightarrow$}\mkern-16.5mu #1}
\def\X{x}

\def\Xpp{x''}
\def\pp{\prime\prime}
\def\p{\prime}

\def\C{{\cal C}}
\newcount\notenumber
\def\clearnotenumber{\notenumber=0}
\def\note{\advance\notenumber by1 \footnote{$^{\the\notenumber}$}}
\clearnotenumber
\font\cs=cmcsc10 scaled\magstep1
\font\bfone=cmbx10 scaled\magstep1
\def\soh#1#2{{\cal G}(#1\vert #2)}
\def\bkb#1#2{\langle{#1}\vert{#2}\rangle}
\def\twa#1{\raise1.5ex\hbox{$\leftrightarrow$}\mkern-16.5mu #1}
\def\tw#1{\tilde{#1}}
\def\bk#1{\vert #1 \rangle}
\def\kb#1{\langle #1 \vert}

\def\u#1{{\bf #1}}
\def\o{\omega}
\def\t{\tau}
\def\D{{\cal D}}
\def\p{\prime}
\def\ve{\varepsilon}
\def\littlesquare#1#2{\vcenter{\hrule width#1in\hbox{\vrule height#2in
   \hskip#1in\vrule height#2in}\hrule width#1in}}
\def\Box{\littlesquare{.0975}{.0975}}

\def\sq{\Box}

%%%%%%%%%%%%% END OF INTERNAL MACROS %%%%%%%%%%%%%%%%

%%%%%%%%%%%%% START OF PAPER %%%%%%%%%%%%%%%%%%%%%%%%

\beginparmode\doublespace
\centerline{\bfone SUM-OVER-HISTORIES ORIGIN }

\centerline{\bfone OF THE COMPOSITION LAWS}

\centerline{\bfone OF RELATIVISTIC QUANTUM MECHANICS}

\centerline{{\bfone AND QUANTUM COSMOLOGY}\footnote{
$^\dagger$}{This work is
supported in part by funds provided by the
U. S. Department of Energy (D.O.E.) under contract
\#DE-AC02-76ER03069.}}\body
\vskip 0.5in
\centerline{{\cs Jonathan J. Halliwell}\footnote{$^\ddagger$}{
j\_halliwell@vax1.physics.imperial.ac.uk\hfil}}
\affil
Center for Theoretical Physics
Laboratory for Nuclear Science
Massachusetts Institute of Technology
Cambridge, MA 02139, USA
and
Blackett Laboratory\footnote{*}{Present~address\hfil}
Imperial College of Science, Technology and Medicine
Prince Consort Road
London SW7 2BZ, UK
\vskip 0.2in
\vskip 0.4in
\author Miguel E. Ortiz\footnote{$^\star$}{ortiz@mitlns.mit.edu\hfil}
\affil
Center for Theoretical Physics
Laboratory for Nuclear Science
Massachusetts Institute of Technology
Cambridge, MA 02139, USA
\vskip 0.6in
\centerline {Submitted to {\sl Physical Review D}}.
\vfill
\noindent{CTP \# 2134\hfil October 1992}\par
\noindent{Imperial-TP/92-93/06\hfil $\,$}\par
\noindent{gr-qc/9211004\hfil $\,$}
\endpage
\vskip 3pt plus 0.3fill \beginparmode
\singlespace \narrower \centerline{\bf ABSTRACT:}
\par

This paper is concerned with the question of the existence of composition
laws in the sum-over-histories approach to relativistic quantum mechanics
and quantum cosmology,
and its connection with the existence a canonical formulation.
In non-relativistic quantum mechanics, the propagator is represented by a sum
over histories in which the paths move forwards in time. The composition law of
the propagator then follows from the fact that the paths intersect an
intermediate surface of constant time once and only once, and a {\it partition}
of the paths according to their crossing position may be affected. In
relativistic quantum mechanics, by contrast, the propagators (or Green
functions) may be represented by sums over histories in which the paths move
backwards and forwards in time. They therefore intersect surfaces of constant
time more than once, and the relativistic composition law, involving a normal
derivative term, is not readily recovered. The principal technical aim
of this paper is to
show that the relativistic composition law may, in fact, be derived directly
from a sum over histories by partitioning the paths according to their {\it
first} crossing position of an intermediate surface. We review the various
Green functions of the Klein-Gordon equation, and derive their composition
laws. We obtain path integral representations for all Green functions except
the causal one. We use the proper time representation, in which the path
integral has the form of a non-relativistic sum over histories but integrated
over time. The question of deriving the composition laws therefore reduces to
the question of factoring the propagators of non-relativistic quantum mechanics
across an arbitrary surface in configuration space. This may be achieved using
a known result called the Path Decomposition Expansion (PDX). We give a proof
of the PDX using a spacetime lattice definition of the Euclidean propagator. We
use the PDX to derive the composition laws of relativistic quantum mechanics
from the sum over histories. We also derive canonical representations of all of
the Green functions of relativistic quantum mechanics, {\it i.e.}, express them
in the form $\langle x'' | x' \rangle $, where the $\{ | x \rangle \}$ are a
complete set of configuration space eigenstates. These representations make it
clear why the Hadamard Green function $G^{(1)}$ does not obey a standard
composition law. They also give a hint as to why the causal Green function does
not appear to possess a sum over histories representation. We discuss the
broader implications of our methods and results for quantum cosmology, and
parameterized theories generally. We show that there is a close parallel
between the existence of a composition law and the existence of a canonical
formulation, in that both are dependent on the presence of a time-like
Killing vector. We also show why certain naive
composition laws that have been proposed in the past for quantum cosmology are
incorrect. Our results suggest that the propagation amplitude between
three--metrics in quantum cosmology, as constructed from the
sum-over-histories, does not obey a composition law.

\endtitlepage

\head {\bf 1. Introduction}
\taghead{1.}

Quantum theory, in both its development and applications, involves
two strikingly different sets of mathematical tools. On the one hand
there is the canonical approach, involving operators, states,
Hilbert spaces and Hamiltonians. On the other, there is the path
integral, involving sums over sets of histories. For most purposes,
the distinction between these two methods is largely
regarded as a matter of mathematical rigour or calculational
convenience. There may, however, be a more fundamental distinction:
one method could be more general than the other. If this is the case,
then it is of particular interest to explore the connections between
the two formulations, and discover the conditions under which
a route from one method to the other can or cannot be found.

A particular context in which the possible distinction between these
two quantization methods will be important is quantum cosmology.
There, the canonical formulation suffers from a serious obstruction
known as  the ``problem of time'' [\cite{ish,kuch}].  This is the
problem that general relativity does not obviously supply the
preferred time parameter so central to the formulation and
interpretation of quantum theory. By contrast, in sum-over-histories
formulations of quantum theory, the central notion is that of a
quantum-mechanical history. The notion of time does not obviously
enter in an essential way.  Sum-over-histories formulations of
quantum cosmology have therefore been promoted as promising
candidates for a quantum theory of spacetime, because the problem of
time is not as immediate or central, and may even be sidestepped
completely [\cite{harnew}]. In particular, as suggested by Hartle, a
sum-over-histories formulation could exist even though a canonical
formulation may not [\cite{harnew}]. The broad aim of this paper is to
explore this suggestion.

An object that one would expect to play an important role in
sum-over-histories formulations of quantum cosmology is the
``propagation amplitude'' between three-metrics. Formally, it is
given by a functional integral expression of the form,
$$
G(h^{\pp}_{ij},h^{\prime}_{ij})
= \int {\cal D} g_{\mu\nu} \exp
\left( i S[g_{\mu\nu}] \right)
$$
Here, $S[g_{\mu\nu}]$ is the
gravitational action, and the sum is over a class of four-metrics
matching the prescribed three-metrics $h^{\pp}_{ij}$,
$h^{\prime}_{ij}$ on final and initial surfaces. The level of the
present discussion is rather formal, so we will not go into the
details of how such an expression is constructed  (see
Ref. [\cite{hh}] for details), nor shall we address the
important question of its interpretation. It is, however, important
for present purposes to assume that a definition of the sum over
histories exists that is not dependent on the canonical formalism.

The above expression is closely analogous in its construction to the
sum-over-histories representations of the propagators (or Green
functions) of relativistic quantum mechanics, ${\cal
G}(x^{\pp}|x')$, where $x$ denotes a spacetime coordinate. We shall
make heavy use of this analogy in this paper.

In relativistic quantum mechanics, there exist both
sum-over-histories and canonical formulations of the one-particle
quantum theory. In the canonical formulation, one may
introduce a complete set of configuration space states, $ \{ | x \ra
\} $. The propagators may then be shown to
possess canonical representations, {\it i.e.}, they may be expressed
in the form,
$$
{\cal G}(x^{\pp}|x') =
\la x^{\pp} | x' \ra
$$
where the right-hand side denotes a genuine Hilbert space inner product.
By insertion of a resolution of the identity, it may then be shown
that the propagator satisfies a composition law, typically of the form
$$
\la x^{\pp} | x' \ra = \int d \sigma^{\mu}
\ \la x^{\pp} | x \ra  \ {\partial_\mu} \la x | x' \ra
$$
where $d \sigma^{\mu} $ denotes a normal surface element. The
details of this type of construction will be given in later
sections. For the moment, the point to stress is that the existence
of a composition law is generally closely tied to the existence of
canonical representations of $ {\cal G} (x^{\pp}|x') $.

Now in a sum-over-histories formulation of quantum cosmology, the
path integral representation of $ G(h^{\pp}_{ij},h^{\prime}_{ij}) $
is taken to be the starting point. Relations such as the composition
law, characteristic of canonical formulations, cannot be assumed but
hold only if they can be {\it derived} directly from the sum over
histories alone, without recourse to a canonical formulation. In
particular, since the existence of a composition law seems to be a
general feature of the canonical formalism, it is very reasonable to
suppose that the existence of a composition law for a  $
G(h^{\pp}_{ij},h^{\prime}_{ij}) $ generated by the sum over
histories is a {\it necessary condition} for the existence of an
equivalent canonical formulation.

The object of this paper is to determine how a derivation of the
composition law from the sum over histories may be carried out. We
may then ask how this derivation might fail, {\it i.e.}, whether the
necessary condition for the recovery of a canonical formulation of
quantum cosmology from a sum-over-histories formulation is fulfilled.

Of course, a full quantum theory of cosmology, even if it existed,
would be exceedingly complicated. Like many authors declaring
interest in quantum cosmology, therefore, we will focus on the
technically simpler case of the relativistic particle.  As
stated above relativistic quantum mechanics
possesses many of the essential features of quantum
cosmology. Remarks on quantum cosmology of a more general and
speculative nature will be saved until the end. We shall show how
the composition laws of relativistic propagators may be derived
directly from their sum-over-histories representations. To the best
of our knowledge, this derivation has not been given previously. It
is therefore of interest in the limited context of relativistic
quantum mechanics, as well as being a model for the more difficult
problem of quantum cosmology  outlined above.

\subhead {\bf 1(A). The Problem}

In non-relativistic quantum mechanics the propagator,
$ \la \xpp, \tpp | \x',t' \ra $, plays a useful and important role.
It is defined to be the object which satisfies
the Schr\"odinger equation with respect to each argument,
$$
	\left( i { \partial \over \partial t^{\pp}} - \hat H^{\pp}
		\right) \la \xpp, \tpp | \x',t' \ra = 0,
	\eqno(1.1)
$$
(and similarly for the initial point), subject to the initial condition
$$
	\la \xpp, t' | \x',t' \ra = \d^{(n)} ( \xpp - \x' )\ .
	\eqno(1.2)
$$
It determines the solution to the Schr\"odinger equation at time
$\tpp$, given initial data at time $t'$:
$$
	\Psi(\xpp, \tpp)
		= \int d^n \x' \ \bkb{\xpp,\tpp}{\x',t'}\ \Psi( \x', t')\ .
	\eqno(1.3)
$$
{}From this follows the composition law (semigroup property),
$$
	\la \xpp, \tpp | \x',t' \ra = \int d^n \x
		\ \bkb{\xpp,\tpp}{\x,t}\ \bkb{\x,t}{\x',t'}\ .
	\eqno(1.4)
$$

In relativistic quantum mechanics, the most closely analogous object is at
first sight the
causal propagator, $G(\Xpp|\X')$. It is defined to be the object satisfying the
Klein-Gordon equation with respect to each argument,
$$
	\left ( \Box_{x^{\pp}}+ m^2 \right) G(\Xpp|\X') = 0\ ,
	\eqno(1.5)
$$
(and similarly for the initial point) and obeying the boundary conditions
$$
\eqalign{
	\left.G (\x^{\pp}, {x^0}^{\pp} |
		\x^{\prime}, {x^0}^{\prime} )
		\right\vert_{{x^0}^{\pp}={x^0}^\p}&= 0\ ,
	\cr
	{\partial\over\partial {x^0}^{\pp}} \left.
		G(\x^{\pp}, {x^0}^{\pp} | \x^{\prime},
		{x^0}^{\prime} ) \right\vert_{{x^0}^{\pp}={x^0}^\p}& =
		-\d^{(3)} (\x^{\pp} -\x^{\prime} )\ .
	\cr
}
$$
It vanishes outside the lightcone. It determines the solution at a spacetime
point $\Xpp$, given initial data on the spacelike surface $\Sigma$
$$
	\phi(x'')=-\int_\Sigma
	d\sigma^\mu \ G(x''|x')\twa{\partial_\mu}\phi(x')
	\eqno(1.8)
$$
where
$$
	\twa{\partial_\mu}=\ura{\partial_\mu}-\ula{\partial_\mu}\ ,
	\eqno(1.9)
$$
and $d\s^\mu$ is normal to the surface $\Sigma$ in
the future timelike direction.
{}From \(1.8) follows the composition law,
$$
	G(\Xpp|\X') = -\int_\Sigma d\sigma^\mu \ G(\Xpp|x)
		\ \twa{\partial_\mu}
		\ G(x|\X') \ .
	\eqno(1.10)
$$
There are of course a number of other Green functions associated with the
Klein-Gordon equation, and many of them also obey composition laws similar
to \(1.10), involving the derivative operator \(1.9) characteristic of
relativistic field theories. For example, the Feynman Green function obeys
a slightly modified version of \(1.10).

Because of the presence of the derivative operator \(1.9) in \(1.10), the
relativistic and non-relativistic composition laws assume a somewhat different
form. The difference is readily understood. The wave functions of
non-relativistic quantum mechanics obey a parabolic equation, and so are
uniquely determined by the value of the wave function on some initial surface.
By contrast, the wave functions in the relativistic case obey a hyperbolic
equation, so are uniquely determined by the value of the wave function and its
normal derivative on some initial surface, hence the derivative term in
\(1.10).

A convenient way of representing the propagator in non-relativistic quantum
mechanics is in terms of a sum-over-histories. Formally, one writes
$$
	\la \xpp, \tpp | \x',t' \ra  = \sum_{p(\x', t'  \ria \xpp, \tpp)}
		\ \exp \left[i S (\x', t'  \ria \xpp, \tpp ) \right]\ .
	\eqno(1.11)
$$
Here, $ p(\x', t' \ria \xpp, \tpp ) $ denotes the set of paths beginning at
$\x'$ at time $t'$ and ending at $\xpp$ at $\tpp$,
and $ S (\x', t' \rightarrow \xpp, \tpp ) $ denotes
the action of each individual such path. The propagator of non-relativistic
quantum mechanics is obtained by restricting to paths $\x(t)$ that
are single-valued functions of $t$, that is, they
{\it move forwards in time}. There
are many ways of defining a formal object like \(1.11). A common method worth
keeping in mind is the time-slicing definition, in which the time interval is
divided into $N$ equal parts of size $\epsilon$, $N \epsilon = (\tpp -t') $,
and one writes
$$
	\la \xpp, \tpp | \x',t' \ra  =
		 \lim_{N \to \infty} \ \prod_{k=1}^N \ \int
		{ d^n\x_k \over (2 \pi i \epsilon)^{n \over 2} } \ \exp \left[
		i S( \x_{k+1}, t_{k+1}| \x_k, t_k ) \right]\ .
%	\eqno(1.12)
$$
Here, $ \x_0 = \x'$, $t_0 = t'$, $\x_{N+1} = \xpp $, $t_{N+1} = \tpp$ and
$S( \x_{k+1}, t_{k+1}| \x_k, t_k )$ is the action of the classical path
connecting $(\x_k,t_k)$ to $(\x_{k+1},t_{k+1})$. More rigorous definitions also
exist, such as that in
which (the Euclidean version of) \(1.11) is defined as the
continuum limit of a sum over paths on a discrete spacetime lattice. Indeed, we
will find it necessary to resort to such a rigorous definition below.

Given the representation \(1.11) of the propagator, it becomes pertinent to ask
whether the composition law \(1.4) may be derived directly from the
sum-over-histories representation, \(1.11). This is indeed possible. The
crucial
notion permitting such a derivation is that of an exhaustive {\it partition}
of the histories into mutually exclusive alternatives. For consider the surface
labeled by $t$, where $t' \le t \le \tpp $. Because the paths move forwards in
time, each path intersects this surface once and only once, at some point
$\x_t$, say. The paths may therefore be exhaustively partitioned into mutually
exclusive sets, according to the value of $\x$ at which they intersect the
surface labeled by $t$ (see Fig. 1). We write this as
$$
\eqalign{
	&p( \x',t' \ria \xpp,\tpp )
		= \bigcup_{\x_t} p (\x',t' \ria \x_t, t
		\ria \xpp, \tpp)
	\cr
	& p (\x',t' \ria \x_t, t \ria \xpp, \tpp )
		\cap p (\x',t' \ria \y_t,t
		\ria \xpp, \tpp ) = \emptyset
		\qquad{\rm if } \qquad\x_t \ne \y_t \ .
	\cr
}
$$
Each path from $ (\x',t') $ to $(\xpp, \tpp )$ may then be uniquely expressed
as the composition of a path from $(\x',t')$ to $(\x_t,t)$ with a path from
$(\x_t,t)$ to $(\xpp,\tpp)$.
Consider what this implies for the sum-over-histories. First of all,
any sensible definition of the measure in the sum-over-histories
should satisfy
$$
\eqalignno{
	\sum_{p(\x',t' \ria \xpp,\tpp)} &= \sum_{\x_t}
		\ \sum_{p(\x',t' \ria \x_t,t  \ria \xpp,\tpp)}
	\cr
	&= \sum_{\x_t} \ \sum_{p(\x',t' \ria \x_t,t)}
		\ \sum_{p(\x_t, t \ria \xpp,\tpp)}\ .
	&(1.15)
	\cr
}
$$
This is readily shown to be true of the time-slicing definition, for
example. Secondly, the action should satisfy
$$
	S(\x',t' \ria \xpp,\tpp)
		= S(\x',t' \ria \x_t,t) + S(\x_t,t \ria \xpp,\tpp)\ .
	\eqno(1.16)
$$
Combining \(1.15) and \(1.16), it is readily seen that one has,
$$
\eqalignno{
	\la \xpp, \tpp | \x',t' \ra  &=
		\sum_{\x_t} \ \sum_{p(\x',t' \ria \x_t,t)}
		\ \sum_{p(\x_t,t \ria \xpp,\tpp)}
		\ \exp\left[ iS(\x',t' \ria \x_t,t)
		+ iS(\x_t,t \ria \xpp,\tpp) \right]
	\cr
	&= \sum_{\x_t}
		\ \la \xpp, \tpp | \x_t,t \ra  \ \la \x_t, t | \x',t' \ra
		\ .
	&(1.17)
	\cr
}
$$
The composition law therefore follows directly from the partitioning
of the sets of paths in the sum over histories.

Turn now to the relativistic particle. There also, certain Green functions
may be represented by sums-over-histories. Formally, one writes
$$
	{\G} (\Xpp|\X') = \sum_{p(\X' \ria \Xpp )} \ \exp \left[i
		S( \X' \ria \Xpp  ) \right]
	\eqno(1.18)
$$
(we will be precise later about which Green function ${\cal G}$ may be).
In fact, a number of such representations are available, since the
classical relativistic particle is a constrained system, and there is
more than one way of constructing the path integral for constrained
systems [\cite{G}].
Here we shall be largely concerned with those constructions for
which the set of paths summed over in \(1.18) is all paths in spacetime.
In particular, unlike the non-relativistic case, the paths will
generally move {\it forwards and backwards} in the time coordinate, $\X^0$
(see Fig. 2).

It again becomes pertinent to ask whether a
composition law of the form \(1.10) may be derived from the sum-over-histories
representation. However, because the paths move both backwards and forwards in
time, they typically intersect an intermediate surface of constant $\X^0$ many
times, and the points at which they intersect the intermediate surface
therefore {\it do not} affect a partition of the paths into exclusive sets. The
argument for the non-relativistic case, therefore, cannot be carried over
directly to the relativistic case. Furthermore, even if this partition did
work, it would then not be clear how the derivative term in the composition law
might arise from the path representation \(1.18). We are thus led to the
question, is there a different way of partitioning the paths, that leads to a
composition law of the form \(1.10), and explains the appearance of the
derivative term? This question is the topic of this paper.

In detail, we will study sum-over-histories expressions of the form \(1.18) for
relativistic Green functions. We will focus on the ``proper time''
sum-over-histories, in which the Green functions are represented by an
expression of the form
$$
	{\cal G} (\Xpp|\X') = \int dT \ g( \Xpp, T | \X', 0)\ .
	\eqno(1.19)
$$
Here $ g( \Xpp, T | \X', 0) $ is a Schr\"odinger propagator satisfying
\(1.1), \(1.2) and \(1.4), with the Hamiltonian taken to be the Klein-Gordon
operator in \(1.5). $g$ may therefore be represented by a sum over paths
of the form \(1.11). We will derive \(1.19) below, but for the moment note
that \(1.19) will be a solution to \(1.5) if $T$ is taken to have an infinite
range, and will satisfy \(1.5) but with a delta-function on the right-hand
side if $T$ is taken to have a half-infinite range.

\subhead {\bf 1(B). Outline}

We begin in Section 2 by reviewing the various Green functions associated with
the Klein-Gordon equation and their properties. We determine which Green
functions satisfy a composition law of the form \(1.8). We briefly
describe the sum over histories representation, and derive \(1.19).
An important question we address is that of which Green functions are
obtained by the sum over paths \(1.19). We also discuss the connection
of sum over histories representations with canonical representations. By
this we mean representations in which the propagators may be expressed in
the form $\bkb{x}{x'}$, where the $\{\bk{x}\}$ with a single time argument
$x^0$ form a complete set of configuration space eigenstates.

In the representation \(1.19), the time coordinate $x^0$ is treated as a
``spatial'' coordinate, when $g$ is thought of as an ordinary Schr\"odinger
propagator like that of non-relativistic quantum mechanics. Comparing \(1.19)
with the expression to be derived from it, \(1.10), we therefore see that
our problem of factoring the sum over histories \(1.19) across a surface of
constant $\X^0$ is very closely related to that of factoring the sum over
histories \(1.11) not across a surface of constant parameter time $t$, as
in \(1.4), but across a surface on which one of the spatial coordinates is
constant. It turns out that a solution to this problem exists, and the
result goes by the name of the Path Decomposition Expansion (PDX)
[\cite{AK}]. The crucial observation that leads to this result is that
although the paths may cross the factoring surface many times, they may
nevertheless be partitioned into exclusive sets according to the parameter
time and spatial location of their
{\it first crossing} of the surface. We describe this result in Section 3,
and give a rigorous derivation of it.

In Section 4, we give our main result. This is to show how the composition
law \(1.10) follows from the sum over histories \(1.19), using the PDX. We also
explain why certain naive composition laws that have been proposed in the
past are problematic.

Our principal result is admittedly simple, and has been derived largely by
straightforward application of the PDX. However, it has broader significance in
the context of the sum-over-histories approach to quantum theory. In
particular, it is closely related to the question of the conditions under which
a sum-over-histories formulation of quantum theory implies the existence of a
Hilbert space formulation. In Section 5, we therefore discuss the
generalizations and broader implications of our result.

\head{\bf 2. The Propagators of Relativistic Quantum Mechanics}
\taghead{2.}

\subhead{\bf 2(A). Green Functions of the Klein-Gordon Equation}

We begin this section with a review of the various Green functions of the
Klein-Gordon equation in Minkowski space relevant to our discussion.
The section is intended to set
out the conventions we shall use throughout this paper, and to list the
relevant properties of the Green functions.
A metric of signature $(+,-,-,-)$ is used throughout. Readers familiar with
the intricacies of this subject may wish to move directly to section B.

The kernel, $\soh{x}{y}$ of the operator $(\sq+m^2)$ satisfying
$$
	\left(\sq_x+m^2\right)\soh{x}{y}=-\delta^4(x-y)
	\eqno(2.1)
$$
where $x$ and $y$ are four--vectors,
may be shown by Fourier transformation to be given by the expression
$$
	\soh{x}{y}={1\over (2\pi)^4}\int{d^4k}{e^{-ik\cdot (x-y)}\over k^2-m^2}.
	\eqno(2.2)
$$
$\soh{x}{y}$ is not uniquely defined in Minkowski space due to the
presence of poles in the integrand. The $k_0$ integration
$$
	\int^{\infty}_{-\infty} dk_0{e^{ik_0(x^0-y^0)}\over
		{k_0}^2-\u{k}^2-m^2}
$$
has
poles on the real axis at $k_0=\pm(\u{k}^2+m^2)^{1/2}$, and the various
possible deformations of this contour determine the possible solutions to
\(2.1), each with different support properties. Below we shall list some
possible contours and their corresponding Green functions.
Closed contours yield solutions to the Klein-Gordon equation. We
also discuss these below since they play an important role in
relativistic quantum mechanics.
\medskip

\noindent{\bf Wightman functions: $G^+(x\vert y)$ and $G^-(x\vert y)$}
\nobreak\smallskip\nobreak
A closed anti-clockwise contour
around one or other of the poles yields the two Wightman functions
$\pm iG^\pm(x\vert y)$, which are solutions of the Klein-Gordon equation,
and of its positive and negative square roots respectively
$$
	\left[i{\partial\over\partial x^0}
		\mp\left(m^2-\nabla^2_\u{x}\right)^{1/2}\right]
		G^{\pm}(x\vert y)=0.
$$
They are given by
$$
	G^{\pm}(x\vert y)={1\over (2\pi)^3}\int d^4k \theta(k^0)\delta(k^2-m^2)
		e^{\mp ik\cdot (x-y)}
$$
or
$$
	G^{\pm}(x\vert y)={1\over (2\pi)^3}\int_{k_0=\pm\o_\u{k}}
		{d^3\u{k}\over 2\o_\u{k}}e^{-ik\cdot (x-y)}
$$
and are related by
$$
	G^+(x\vert y)=G^-(y\vert x).
$$
The two Wightman functions satisfy relativistic composition laws
$$
	G^{\pm}(x''\vert x')=\pm i \int_\Sigma d\sigma^\mu
		G^{\pm}(x''\vert x)
		\twa{\partial_\mu} G^{\pm}(x\vert x')
	\eqno(2.60)
$$
(where $d\s^\mu$ is normal to $\Sigma$ and future pointing)
and are orthogonal in the sense that
$$
	\int_\Sigma d\sigma^\mu
		G^{\pm}(x''\vert x)\twa{\partial_\mu}G^{\mp}(x\vert x')=0.
$$
In field theory they are given by the expressions
$$
	G^+(x\vert y)=\kb{0}\phi(x)\phi(y)\bk{0}
$$
and
$$
	G^-(x\vert y)=\kb{0}\phi(y)\phi(x)\bk{0}\ .
$$
\medskip

\noindent{\bf Feynman propagator: $G_F(x\vert y)$}
\nobreak\smallskip\nobreak
A contour going under the left pole and above the right gives the
Feynman propagator. This satisfies equation \(2.1), and may be written as
$$
	iG_F(x\vert y)=\theta(x^0-y^0)
		G^+(x|y)+\theta(y^0-x^0)G^-(x|y).
$$
Alternatively,
$$
\eqalign{
        G_F(x\vert y)&={-i\over (2\pi)^4}\int^\infty_{0}dT\int{d^4 k}
                e^{-i[k\cdot (x-y)-T(k^2-m^2+i\varepsilon)]}
	\cr
	&={1\over (2\pi)^4}\int d^4k{e^{-ik\cdot (x-y)}\over k^2-m^2+i\ve}.
	}
$$
It may be checked that $G_F$ obeys a relativistic composition law
$$
	G_F(x''\vert x')=-\int_\Sigma
		d\sigma G_F(x''\vert x)\twa{\partial_n}
		G_F(x\vert x')\ ,
	\eqno(2.300)
$$
where $\Sigma$ is an arbitrary spacelike 3--surface, and $\partial_n=
n^\mu\partial_\mu$ with $n^\mu$ now
the normal to $\Sigma$ in the direction of propagation.
In free scalar field theory, the Feynman propagator is of course given by
$$
      iG_F(x\vert y)=\kb{0}T\left(\phi(x)\phi(y)\right)\bk{0}.
$$

\medskip
\noindent{\bf Causal Green function: $G(x\vert y)$}
\nobreak\smallskip\nobreak
A closed clockwise
contour around both poles gives what is generally known as the
commutator or causal
Green function, which is written simply as $G(x\vert y)$. It is a
solution of the Klein-Gordon equation.
$G(x\vert y)$ has the following representations
$$
	G(x\vert y)={-i\over (2\pi)^3}\int {d^4{k}}\varepsilon(k^0)
          \delta(k^2-m^2)e^{-ik\cdot (x-y)}
	\eqno(2.68)
$$
or
$$
	G(x\vert y)={-1\over (2\pi)^3}\int {d^3\u{k}\over \o_\u{k}}
		\sin\left[\o_\u{k}(x^0-y^0)\right]e^{i\u{k}\cdot(\u{x-y})}.
$$

Since
$$
      \left.G(\u{x},x^0\vert \u{y},y^0)\right\vert_{x^0=y^0}=0,\qquad
		\left.{\partial\over\partial x^0}G(\u{x},x^0\vert\u{y},y^0)
		\right\vert_{x^0=y^0}=-\delta^3(\u{x-y}),
$$
and $G$ is Lorentz invariant,
it has support only within the light cone of $x-y$.
$G$ also obeys the relativistic composition law
$$
	G(x''\vert x')=-\int_\Sigma d\sigma^\mu
		G(x''\vert x)\twa{\partial_\mu} G(x\vert x').
$$
and, as mentioned in the Introduction,
it propagates solutions $\phi(x)$ of the Klein-Gordon equation via
$$
	\phi(y)=-\int_\Sigma d\sigma^\mu
		G(y\vert x)\twa{\partial_\mu}\phi(x).
$$
In field theory, $G$ is given by the commutator
$$
	iG(x\vert y)=\kb{0}\left[\phi(x),\phi(y)\right]\bk{0}=
		\left[\phi(x),\phi(y)\right].
$$
Finally, note that
$$
      iG(x\vert y)=G^+(x\vert y)-G^-(x\vert y)\ .
$$

\medskip
\noindent{\bf Hadamard Green function: $G^{(1)}(x\vert y)$}
\nobreak\smallskip\nobreak
A closed figure of eight contour around the two poles gives the Hadamard or
Schwinger Green function $iG^{(1)}(x\vert y)$,
which is a solution of the Klein-Gordon
equation. It may be written as
$$
\eqalign{
	G^{(1)}(x\vert y)&={1\over (2\pi)^3}\int {d^4{k}}
          \delta(k^2-m^2)e^{-ik\cdot (x-y)}
        \cr
        &={1\over (2\pi)^4}\int^\infty_{-\infty}dT\int{d^4 k}
          e^{-i[k\cdot (x-y)-T(k^2-m^2)]}
        }
	\eqno(2.3)
$$
or
$$
	G^{(1)}(x\vert y)={1\over (2\pi)^3}\int {d^3\u{k}\over \o_\u{k}}
		\cos\left[\o_\u{k}(x^0-y^0)\right]e^{i\u{k}\cdot(\u{x-y})}.
$$
Perhaps the most important property of $G^{(1)}(x\vert y)$ is that it does not
satisfy the standard relativistic composition law. In fact
$$
\eqalign{
	G^{(1)}(x''\vert x')=&-\int_\Sigma
		d\sigma^\mu
		G(x''\vert x)\twa{\partial_\mu} G^{(1)}(x\vert x')
	\cr
	=&-\int_\Sigma
		d\sigma^\mu G^{(1)}(x''\vert x)\twa{\partial_\mu}G(x\vert x')
	}
	\eqno(2.110)
$$
and
$$
	G(x''\vert x')=\int_\Sigma d\sigma^\mu
		G^{(1)}(x''\vert x)\twa{\partial_\mu} G^{(1)}(x\vert x').
$$
In field theory, $G^{(1)}$ is given by the anti-commutator
$$
	G^{(1)}(x\vert y)=\kb{0}\left\{\phi(x),\phi(y)\right\}\bk{0}.
$$
It is related to the Wightman functions via
$$
        G^{(1)}(x\vert y)=G^+(x\vert y)+G^-(x\vert y).
$$
\medskip

\noindent{\bf Newton--Wigner propagator: $G_{NW}(\u{x},x^0\vert \u{y},y^0)$}
\nobreak\smallskip\nobreak
The Newton--Wigner propagator is a solution of the Klein-Gordon equation, and
indeed of its first order positive square root.
It is not given by the integral \(2.2) for any contour. We nevertheless
include it since it plays an important role in the quantum mechanics of the
relativistic particle. $G_{NW}$ is defined by
$$
	G_{NW}(\u{x},x^0\vert\u{y},y^0)={1\over (2\pi)^3}\int_{k_0=\o_\u{k}}
		{d^3\u{k}}e^{-ik\cdot (x-y)}\ .
	\eqno(2.4)
$$
The support property
$$
	\left.G_{NW}(\u{x},x^0\vert\u{y},y^0)\right\vert_{x^0=y^0}
		=\delta^3(\u{x-y})
$$
shows that $G_{NW}$
is analogous to the quantum mechanical propagator
\(1.2). It propagates solutions of the first order Schr\"odinger
equation with Hamiltonian $H=(k^2-m^2)^{1/2}$. $G_{NW}$
also obeys the usual quantum mechanical composition law
$$
	G_{NW}(\u{x''},{x^0}''\vert\u{x},x^0)=\int d^3\u{x'}G_{NW}(\u{x''},
		{x^0}''\vert
		\u{x'},{x^0}')G_{NW}(\u{x'},{x^0}'\vert\u{x},x^0)\ .
	\eqno(2.70)
$$
Finally, note that $G_{NW}$
is not Lorentz invariant.

An analogous operator, which we shall call the negative frequency
Newton--Wigner propagator, may also be defined. It is given by
$$
	\tw{G}_{{NW}}(\u{x},x^0\vert\u{y},y^0)=
		{1\over (2\pi)^3}
		\int_{k_0=-\o_\u{k}} d^3\u{k}e^{-ik\cdot (x-y)}
$$
and
has the same support properties as $G_{NW}(\u{x},x^0\vert\u{y},y^0)$.
It solves the negative
frequency square root of the Klein-Gordon equation and propagates its
solutions.

\subhead{\bf 2(B). Sum Over Histories Formulation of
Relativistic Quantum Mechanics}

We are interested in Green functions which may be represented by sums over
histories of the form \(1.18). We will take sum-over-histories expressions of
the form \(1.18) as our starting point and determine which Green functions they
give rise to. The expression \(1.18) is rather formal as it stands, and various
aspects of it need to be specified more precisely before it is properly and
uniquely defined. These include the action, class of paths, gauge-fixing
conditions, and the domains of integration of certain variables. The particular
Green function obtained will depend on how these particular features are
specified. We note, however, that there is no guarantee that {\it all}
known Green functions may be obtained in this way, and indeed, we are not
able to obtain the causal propagator, $G$.

The action for a relativistic particle is usually written as
$$
	S=-m \ \int^{\t^{\p\p}}_{\t^\p}d\t \ \left[{\partial x^\mu\over
		\partial\t}{\partial x^\nu\over
		\partial\t}\eta_{\mu\nu}\right]^{1/2},
	\eqno(2.5)
$$
the length of the worldline of the particle in Minkowski space.
$\t$ parameterizes the worldline, and $S$ is invariant under
reparameterizations $\t\to f(\t)$. Since \(2.5) is highly non-linear, its
quantization presents certain difficulties which have hitherto prevented
its direct use in a sum over histories. These difficulties may be
bypassed by the introduction of an auxiliary variable $N$, which can be
thought of as a metric on the particle worldline. The action may then be
rewritten as
$$
	S=-\int^{\t^{\p\p}}_{\t^\p}d\t \left[{\dot{x}^2\over 4 N}+
                                        m^2 N\right]
$$
where a dot denotes a derivative with respect to $\t$. Passing to a
Hamiltonian form, the action becomes
$$
	S=\int^{\t^{\p\p}}_{\t^\p}d\t\left[p_\mu \dot x^{\mu}
		+N H\right]
	\eqno(2.6)
$$
where $N$ is now a Lagrange multiplier which enforces the constraint
$H=(p^2-m^2)=0$. The Hamiltonian form \(2.6) of the action is still
invariant under repara\-meter\-izations. Infinitesimally, these
are generated by the constraint $H$,
$$
	\delta x=\ve(\t)\{x,H\},\qquad \delta p=\ve(\t)\{p,H\},\qquad
		\delta N=\dot\ve(\t)
$$
for some arbitrary parameter $\ve(\t)$.
Since $H$ is quadratic in momentum, the action is only invariant
up to a surface term [\cite{HT}]
$$
	\delta S=\left[\ve(\t)\left(p{\partial H\over \partial
		p}-H\right)\right]^{\t''}_{\t'}
$$
which constrains the reparameterizations at the end points to obey
$$
	\ve(\t'')= 0= \ve(\t').
$$

We shall discuss briefly the use of a sum over histories to evaluate the
amplitude for a transition from $x'$ to $x''$, which we shall write as
$\soh{x''}{x'}$. The sum is
over paths beginning at $x'$ at parameter time $\t'$,
and ending at $x''$ at parameter time $\t''$. Trajectories may in
principle move forwards and backwards in the physical
time $x^0$, although it is
also possible to define an amplitude constructed from paths that move only
forwards in $x^0$, as we shall discuss below.

It is necessary to fix the reparameterization invariance, and this may be done
in a number of ways. We shall give a brief description of the two most commonly
used prescriptions: the so-called proper time gauge $\dot N = 0$, and the
canonical gauge $x^0=\t$. The proper time gauge is a good prescription in the
Gribov sense [\cite{G}]. The canonical gauge has the feature that it restricts
the class of paths in configuration space to move forwards in the time
coordinate $x^0$. These two gauge-fixing conditions lead to quite different
results. There is, however, no conflict with the standard result that the path
integral is independent of the choice of gauge-fixing [\cite{BFV}]. That result
applies only to families of gauge-fixing conditions which may be smoothly
deformed into each other, which is not true of the two gauges described above.

\bigskip
\noindent{\bf Proper Time Gauge $\dot{N}=0$}
\nobreak\medskip\nobreak
The proper time gauge has been extensively discussed in the literature
[\cite{G},\cite{HT},\cite{T},\cite{JJH}],
and we shall therefore only state some well-known results.

The condition $\dot{N}=0$ is implemented by adding a gauge fixing
term $\Pi\dot{N}$ to
the Lagrangian. The BFV prescription also requires the
addition of a ghost term (details may be found in [\cite{JJH}]). The path
integration over the ghosts factorises, and the gauge fixing condition,
realised by the integration over the Lagrange multiplier $\Pi$,
reduces the functional integration over $N$ to a
single integration, leaving
$$
	\soh{x''}{x'}=\int dN(\t''-\t')\int \D p\D x\
		\exp\left[i\int^{\t''}_{\t'}d\t\left[p\dot{x}-NH\right]
		\right].
	\eqno(2.50)
$$
Redefining $T=N(\t''-\t')$, this may be rewritten as
$$
	\soh{x''}{x'}=\int dT g(x'',T\vert x',0)
$$
where $g(x'',T\vert x',0)$ is an ordinary
quantum mechanical transition amplitude with Hamiltonian $H=p^2-m^2$.
The amplitude is given explicitly by
$$
	\soh{x''}{x'}={1\over (2\pi)^4}\int dT \int d^4p \;
		e^{i[p\cdot(x''-x')-T(p^2-m^2)]}.
$$
All that remains is to specify the range of $T$ integration. If $T$ is
integrated over an infinite range, then the Hadamard Green function is
obtained,
$$
	\soh{x''}{x'}=G^{(1)}(x''\vert x')
$$
On the other hand, if the range of integration is limited to
$T\in [0,\infty)$, then, introducing a regulator to make the $T$ integration
converge, the Feynman Green function is obtained,
$$
	\soh{x''}{x'}=iG_F(x''\vert x').
$$
{}From this the sum-over-histories representations of $G^{\pm}$ are readily
obtained. $G^+$ is obtained by taking $T>0$ and ${x^0}'>x^0$, or $T<0$ and
${x^0}'<x^0$, with the reverse yielding $G^-$.
In all of these cases, the class of paths is taken to be all paths in
spacetime connecting the initial and final points. Note that the causal
propagator $G(x''|x')$ is not obtained by these means. We will return to this
point later.

\bigskip
\noindent{\bf Canonical Gauge $x^0=\tau$}
\nobreak\medskip\nobreak
It is also of interest to consider a sum over histories in which the paths
are restricted to move forwards in the physical time $x^0$.
On this class of paths, $x^0=\t$
may be shown to be a valid gauge choice, provided that one sets up the
parameter time interval so that $\t'={x^0}'$ and $\t''={x^0}''$.
It may be implemented in the
action by the addition of a gauge fixing term $\Pi(x^0-\t)$.
An evaluation of the path integral, using an infinite range for $N$,
leads to the amplitude
$$
	\soh{\u{x}'',{x^0}''}{\u{x}',{x^0}'}=
		G_{NW}(\u{x}'',{x^0}''\vert\u{x}',{x^0}')
		+\tw{G}_{NW}(\u{x}'',{x^0}''\vert\u{x}',{x^0}').
	\eqno(2.100)
$$
This includes contributions from both positive and negative frequency
sectors of the relativistic particle, in the sense that trajectories with
both positive and negative $p_0$ are summed over. If the integrations over
$N$ are
restricted to positive $N$ (equivalently, a factor $\theta(N)$ is included
on every time slice), then only the positive frequency sector is included.
In this case the amplitude is given by the
Newton--Wigner propagator
$$
	\soh{x''}{x'}=G_{NW}(x''\vert x').
$$

The choice of a canonical gauge leads in both cases to an amplitude which
is not Lorentz invariant, a consequence of the preferred status
acquired by the co-ordinate $x^0$. A comprehensive discussion
of this gauge may be found in [\cite{HK}].

\subhead{\bf 2(C). Canonical Formulation of
Relativistic Quantum Mechanics}

We have listed the various Green functions, their composition laws, and
their path integral representations, where they exist. In this subsection
we discuss the connection of these considerations with the canonical
quantization of the relativistic particle. In particular, we ask whether
the various Green functions have canonical representations of the form
$ \langle x'' | x' \rangle $, where the $\{ | x' \rangle \} $ are complete
sets of configuration space eigenstates for any particular value of $x^0$,
and are constructed by taking
suitable superpositions of physical states ({\it i.e.} ones satisfying the
constraint). We will find that essentially all of the Green functions may be
so represented. Which Green function is obtained depends on the choice
inner product in the space of physical states, and on which states
are included in the superposition (positive frequency, negative frequency
or both). These considerations will shed some light on various features
of the composition laws.

Dirac quantization of the relativistic particle leads to a space of states
which may be expressed in terms of a complete set of momentum eigenstates
$$
	\hat{p}_{\mu}\bk{p}=p_\mu\bk{p}
$$
subject to the additional constraint
$$
	(p^2-m^2)\bk{p}=0.
$$
The solutions to the constraints may be labelled by the eigenstates of the
3-momentum $\u{p}$, and we denote them $\bk{\u{p}}$. The states
$\bk{\u{p}}$ are not complete, since there remains an ambiguity in the action
of
$$
	\hat{p_0}\bk{\u{p}}=\pm\left(\u{p}^2+m^2\right)^{1/2}\bk{\u{p}}.
$$
For free particles, the positive and negative frequency states decouple.
Canonical representations are therefore possible involving
the positive and negative frequency sectors separately, or both together.
We consider each in turn. Our aim is to find canonical representations
in which each of the Green functions may be represented in the form
$ \langle x''|x' \rangle $.
\bigskip
\noindent{\bf Positive Frequency Sector}
\nobreak\medskip\nobreak
In the positive frequency sector, $p_0>0$, the $\bk{\u{p}}$ such that
$$
	\hat{p_0}\bk{\u{p}}=\left(\u{p}^2+m^2\right)^{1/2}\bk{\u{p}}
$$
form a complete basis. The
appropriate choice of inner product is
$$
	\bkb{\u{p}}{\u{p'}}=2\o_\u{p}\delta(\u{p}-\u{p'})
$$
and the completeness relation
$$
	1=\int {d^3\u{p}\over 2\o_\u{p}} \bk{\u{p}}\kb{\u{p}}
$$
follows,
where $\o_\u{p}=(\u{p}^2+m^2)^{1/2}$.
Two choices of configuration space representations of this Hilbert
space are possible: the Newton--Wigner representation, and the
relativistically invariant representation.

\medskip

\font\cs=cmcsc10
\noindent{\cs Newton--Wigner Representation}
\nobreak\smallskip\nobreak
{}From the basis $\bk{\u{p}}$, we may change to the Newton--Wigner basis
defined by the states
$$
	\bk{\u{x},x^0}={1\over(2\pi)^{3/2}}\int_{p_0=\o_\u{p}}
		{d^3\u{p}\over (2\o_\u{p})^{1/2}}
		e^{ip\cdot x}\bk{\u{p}}.
$$
They are not Lorentz invariant, but
they are orthogonal at equal times and satisfy the completeness relation
$$
	1=\int d^3\u{x}\ \bk{\u{x},x^0}\kb{\u{x},x^0}.
	\eqno(2.7)
$$
Any Newton--Wigner wave function $\Psi(\u{x},x^0)=\bkb{\u{x},x^0}{\Psi}$
satisfies the positive square root of the Klein-Gordon equation,
$$
	i{\partial\over \partial x^0}\Psi=(m^2-\nabla^2)^{1/2}\Psi
$$
which reflects the fact that we are only considering the positive frequency
excitations. The inner product on wave functions is the usual one
$$
	\bkb{\Phi}{\Psi}=\int d^3\u{x}\ \Phi^\dagger(\u{x},x^0)\Psi(\u{x},x^0)
$$
and the propagator
$\bkb{\u{x},x^0}{\u{x'},{x^0}'}$ is precisely the Newton--Wigner propagator
\(2.4). Its composition law \(2.70) follows immediately from \(2.7).

\medskip

\noindent{\cs Relativistic Representation}
\nobreak\smallskip\nobreak
It is possible to define a Lorentz invariant configuration space
representation, using the basis states
$$
	\bk{x}={1\over(2\pi)^{3/2}}\int_{p_0=\o_\u{p}}
		{d^3\u{p}\over 2\o_\u{p}}
		e^{ip\cdot x}\bk{\u{p}}
$$
where a the states $\bk{x}$ with $x^0$ fixed
form a basis on the space of physical states.
They are not orthogonal, since at equal times $x^0$ one has
$$
	\bkb{x}{x'}={1\over (2\pi)^3}\int_{p_o=\o_\u{p}}
		{d^3\u{p}\over 2\o_\u{p}}e^{-i\u{p}\cdot(\x-\x')}.
	\eqno(2.67)
$$
They satisfy the relativistic completeness relation
$$
	1=i\int_\Sigma d\sigma^\mu \bk{x}\twa{\partial_\mu}\kb{x}.
	\eqno(2.8)
$$
where $\Sigma$ is an arbitrary spacelike 3--surface.
The corresponding wave functions $\psi(x)=\bkb{x}{\psi}$ solve the positive
square root of the Klein-Gordon equation and their positive definite inner
product is given by the usual relativistic expression
$$
	\bkb{\phi}{\psi}=i\int_\Sigma d\sigma^\mu\phi^\dagger(x)
		\twa{\partial_\mu}\psi(x).
	\eqno(2.9)
$$
The propagator $\bkb{x'}{x}$ given in Eq. \(2.67) is equal to $G^+(x'\vert x)$.

Similarly, by restricting to the negative frequency sector, it is readily shown
that $\bkb{x'}{x}$ is equal to $G^-(x'\vert x)$. Canonical representations of
$G^{\pm}$ are therefore readily obtained. A canonical representation of the
Feynman Green function comes from those for $G^+$ and $G^-$, although it is not
immediately apparent how to construct a more direct one than this. These
propagators all obey suitably modified versions of
the relativistic composition law \(1.10), as readily follows
from the completeness relation \(2.8).
\bigskip

\noindent{\bf Positive and Negative Frequency Sectors}
\nobreak\medskip\nobreak
The discussion above provides a canonical description of both the Feynman
and New\-ton--Wigner propagators which arose in the path integral formulation
of section 2(B), with $N>0$.
However, if the range of integration of the lapse function
$N$ is not restricted to a half--infinite range for the proper time gauge,
we saw that the path integral leads to the propagator
$ \soh{x''}{x'}=G^{(1)}(x''\vert x') $
where $G^{(1)}$ is the Hadamard Green function \(2.3). Since restricting
$N$ to be positive (or negative) in the sum over histories appears to
correspond to the positive (or negative) frequency sectors in the canonical
representations, it is very plausible that a canonical representation of
$G^{(1)}$ will involve both sectors simultaneously. These is indeed
the case, as we now show.

In momentum space, there are two orthogonal copies of the space of states
$\bk{\u{p}}$. We label these two copies $\bk{\u{p},\pm}$ where
$$
	\hat{p_0}\bk{\u{p},\pm}=\pm(\u{p}^2+m^2)^{1/2}\bk{\u{p},\pm}.
$$
The space of states is now a sum of the two copies, on which we choose
the completeness relation
$$
	1=\int {d^3\u{p}\over 2\o_\u{p}} \left[\bk{\u{p},+}\kb{\u{p},+}+
		\bk{\u{p},-}\kb{\u{p},-}\right].
	\eqno(2.10)
$$
The corresponding inner product is positive definite for all states
$$
	\bkb{p, i}{p',j}=2\o_\u{p}\delta^3(\u{p}-\u{p'})\delta_{ij}
	\eqno(2.11)
$$
where $i,j=\pm$ [\cite{HT}].
\medskip

\noindent{\cs Newton--Wigner Representation}
\nobreak\smallskip\nobreak
We define Newton--Wigner states as
$$
	\bk{\u{x},x^0}=
		\left[
		\int_{p_0=\o_\u{p}} {d^3\u{p}\over (2\o_\u{p})^{1/2}}
		e^{ip\cdot x}\bk{\u{p},+}
		+\int_{p_0=-\o_\u{p}} {d^3\u{p}\over (2\o_\u{p})^{1/2}}
		e^{ip\cdot x}\bk{\u{p},-}\right]
$$
where $p=(\o_\u{p},\u{p})$. This definition
is compatible with \(2.10) and \(2.11) provided that the usual completeness
relation \(2.7) is amended. Defining
$$
	\bk{\u{x},x^0,+}=
		\int_{p_0=\o_\u{p}} {d^3 \u{p}\over (2\o_\u{p})^{1/2}}
		e^{ip\cdot x}\bk{\u{p},+},\qquad
	\bk{\u{x},x^0,-}=
		\int_{p_0=-\o_\u{p}} {d^3 \u{p}\over (2\o_\u{p})^{1/2}}
		e^{ip\cdot x}\bk{\u{p},-},
$$
as the positive and negative frequency parts of $\bk{\u{x},x^0}$, \(2.7) is
replaced by
$$
	1=\int d^3\u{x}\left[\bk{\u{x},x^0,+}\kb{\u{x},x^0,+}+
		\bk{\u{x},x^0,-}\kb{\u{x},x^0,-}\right].
$$
The propagator for Newton--Wigner states is then given by
$$
	\bkb{\u{x},x^0}{\u{x}',{x^0}'}=G_{NW}(\u{x},x^0\vert
		\u{x}',{x^0}')+
		\tw{G}_{NW}(\u{x},x^0\vert\u{x}',{x^0}')\ .
$$
This is precisely the amplitude derived in section 2(B), Eq.\(2.100).

Note that now the wave function $\Psi(\u{x},x^0)=\bkb{\u{x},x^0}{\Psi}$
solves only the second order Klein-Gordon equation.
The inner product on wave functions $\Psi(\u{x},x^0)$ is
$$
	\bkb{\Phi}{\Psi}=\int d^3\u{x}\left[\Phi_+^\dagger
		(\u{x},x^0)\Psi_+(\u{x},x^0)+
		\Phi_-^\dagger(\u{x},x^0)\Psi_-(\u{x},x^0)\right].
$$

\medskip

\noindent{\cs Relativistic Representation}
\nobreak\smallskip\nobreak
Lorentz invariant states in this canonical representation involving
positive and negative frequency states may be defined by
$$
	\bk{x}= \int_{p_0=\o_\u{p}} {d^3\u{p}\over 2\o_\u{p}}
		e^{ip\cdot x}\bk{\u{p},+}
		+\int_{p_0=-\o_\u{p}} {d^3\u{p}\over 2\o_\u{p}}
		e^{ip\cdot x}\bk{\u{p},-},
	\eqno(2.12)
$$
The usual treatment of the relativistic particle involves these states
along with the usual relativistic completeness relation \(2.8), which is
equivalent to the usual relativistic inner product \(2.9) on wave functions
[\cite{Schiff}]. However, it is
well known that \(2.9) is not positive definite on the class of functions
with both positive and negative frequency parts. Hence, \(2.8) and \(2.12) are
not compatible with the positive definite inner product \(2.11). In fact,
working backwards, they imply that
$$
	\bkb{\u{p},\pm}{\u{p'},\pm}=\pm 2\o_\u{p}\delta(\u{p}-\u{p'})\ ,
        \eqno(2.200)
$$
and
$$
	1=\int {d^3\u{p}\over 2\o_\u{p}} \left[\bk{\u{p},+}\kb{\u{p},+}-
		\bk{\u{p},-}\kb{\u{p},-}\right].
$$
If we are to keep \(2.8) and \(2.12), therefore, we must use the indefinite
inner product \(2.200) in place of \(2.11). In this way, we do in fact obtain
the canonical representation for the causal Green function, for it is
readily shown that one has,
$$
	\bkb{x}{x'}=iG(x\vert x').
$$
Its composition law follows from inserting the resolution of the identity,
\(2.8).

A different canonical representation may be obtained by keeping \(2.12)
and the positive definite inner product \(2.11), but modifying the
completeness relation \(2.8). Define
$$
      \bk{x,+}=\int_{p_0=\o_\u{p}}
		{d^3 \u{p}\over 2\o_\u{p}} e^{ip\cdot x}\bk{\u{p},+},\qquad
      \bk{x,-}=\int_{p_0=-\o_\u{p}}
		{d^3 \u{p}\over 2\o_\u{p}} e^{ip\cdot x}\bk{\u{p},-},
$$
and replace \(2.8) by
$$
	1=i\int_\Sigma d\sigma^\mu
		\left[\bk{x,+}\twa{\partial_\mu}\kb{x,+}-
        	\bk{x,-}\twa{\partial_\mu}\kb{x,-}\right]
	\eqno(2.13)
$$
which is compatible with \(2.10) and \(2.11).
The appropriate relativistic propagator is
$$
      \bkb{x}{x'}=G^{(1)}(x\vert x')=G^+(x\vert x')+G^-(x\vert x')\ ,
$$
this giving a canonical representation of the Hadamard Green function.
The wave functions $\psi(x)=\bkb{x}{\psi}$ satisfy
the Klein-Gordon equation and obey
$$
      \psi(x')=i\int_\Sigma d\sigma^\mu \left[G^+(x'\vert
		x)\twa{\partial_\mu}\psi_+(x)-
        	G^-(x'\vert x)\twa{\partial_\mu}\psi_-(x)\right]
$$
where $\psi_\pm$ are the positive and negative frequency parts of $\psi$.
Since
$$
      \int_\Sigma d\sigma^\mu G^+\twa{\partial_\mu}\psi_-
		=\int_\Sigma d\sigma^\mu G^-\twa{\partial_\mu}\psi_+=0,
$$
it follows that
$$
      \psi(x')=-\int_\Sigma d\sigma^\mu G(x',x)\twa{\partial_\mu}\psi(x),
$$
where $iG=G^+-G^-$ is the causal Green function. This is precisely the
evolution equation we expect for
$\psi$ a solution to the Klein-Gordon equation with both positive and negative
frequency parts. The unusual form \(2.13) of the completeness relation
explains how it is
that the Green function $G^{(1)}$, which does not propagate solutions of
the Klein-Gordon equation, is nevertheless
compatible with causal evolution of a
wave function $\psi(x)$.
The inner product on $\psi(x)$ is now not \(2.9) but rather
$$
	\bkb{\phi}{\psi}
		=i\int_\Sigma d\sigma^\mu
		\left[\phi^\dagger_+(x)\twa{\partial_\mu}\psi_+(x)
	        -\phi^\dagger_-(x)\twa{\partial_\mu}\psi_-(x)\right]
	\eqno(2.77)
$$
which is by construction positive definite.

We note that a significant and seemingly anomalous property of
$G^{(1)}$ is that, unlike all the other Green functions, it does not obey a
composition law of the usual form, but instead obeys \(2.110)\note{
That this is somewhat puzzling was, to our knowledge, first
noticed by Ikemori [\cite{J}].}.
Our study of
canonical representations now makes it clear why this is. The composition laws
of $G_F$, $G^{\pm}$ and $G$ readily follow from their canonical representations
$\bkb{x}{x'}$ by simply inserting the resolution of the identity, Eq.
\(2.8).
Recall, however, that the canonical representation of $G^{(1)}$ involves
dropping \(2.8) in favour of \(2.13), from which follows the result,
$$
	G^{(1)}(x''\vert x')
	=i\int_\Sigma d\sigma^\mu
		\left[G^+(x''\vert x)\twa{\partial_\mu}G^+(x\vert x')
        	-G^-(x''\vert x)\twa{\partial_\mu}G^-(x\vert x') \right]
	\eqno(2.80)
$$
which is readily shown to be equivalent to \(2.110). The important point,
therefore, is that the unusual form of the composition law for $G^{(1)}$
is explained by the non-standard resolution of the identity in its
canonical representation, which is in turn necessitated by the assumed
positive definite inner product on both positive and negative frequency
states.

We have therefore derived canonical representations of all the Green functions.
Our results, together with the composition laws and path integral
representations are summarized in Table 1.

Finally, we make the following comments on the connection between
the sum over histories and canonical formulations of relativistic
quantum mechanics. A sum over histories representation of a given
propagator may be derived from its canonical representation, by a
standard procedure, which involves inserting resolutions of the
identity into the canonical expression $ \la x | x' \ra $ (except
for the causal Green function -- see below). It is then reasonable
to ask how one might proceed in the opposite direction, {\it i.e.},
given a propagator, as supplied by the sum over histories,
how does one derive the Hilbert space inner product from which the
canonical representation is constructed? The answer to this question
lies in the observation that the inner products given above
for the relativistic representations all have the form,
$$
\la \phi | \psi \ra = -\int_{\Sigma,\Sigma'} d\sigma^\mu d\sigma^{\mu'}
\ \phi^{\dag}(x) \ \twa{\partial_\mu} \,
{\cal G}(x|x') \ \twa{\partial_{\mu'}}\, \psi(x')
$$
So, for example, by taking ${\cal G}$ to be $G^+$, one obtains the
inner product \(2.9). This observation is a natural starting point for
the possible derivation of a canonical formulation from a sum over
histories, as we shall discuss further in Section 5.

\subhead{\bf 2(D). Summary of Section 2}

In words, our results may be summarized as follows:

\item{(a)} The Green functions $G^{\pm}$ and $G_F$ obey standard composition
laws (Eqs. \(2.60) and \(2.300)).
They may be obtained by sums over histories over either
positive or negative proper time. Their canonical representations may
be obtained by restriction to the positive or negative frequency sector, with a
positive definite inner product and with the usual resolution of the identity.

\item{(b)} The causal Green function $G$ obeys the standard composition law. It
does not obviously have a sum over histories representation. Its canonical
representation involves both the positive and negative frequency sectors, with
an indefinite inner product and the usual resolution of the identity.

\item{(c)} The Hadamard Green function $G^{(1)}$ does not obey the
standard composition law. It may be obtained by a sum over histories
over both positive and negative proper time. Its canonical representation
involves both positive and negative frequency sectors, with a positive
definite inner product and a non-standard resolution of the
identity. The latter explains the absence of the usual composition law.

\item{(d)} The Newton--Wigner propagator obeys the composition law
of the non\--re\-la\-ti\-vis\-tic type. It may be obtained by a sum over
histories of the form \(2.50) in which the paths move forwards in the
physical time $x^0$. It has a canonical  representation in the positive
frequency sector with a positive definite inner product, with the usual
quantum mechanical resolution of the identity. We will find below that an
alternative, rather novel representation in the proper time gauge is also
available.

It is striking that unlike all the other Green functions, the causal
Green function is represented canonically with an indefinite inner product.
We conjecture that this is the reason why it does not have an obvious
sum-over-histories representation in configuration space of the form
\(1.18). Briefly, a phase space path integral representation may be
constructed by inserting resolutions of the identity into the canonical
expression $\bkb{x}{x'}$, and the configuration space path integral is
obtained by integrating out the momenta. For the causal Green function,
however, the indefinite inner product leads to the appearance of factors of
$\varepsilon(p_0)$ in the phase space path integral ({\it c.f.} Eq. \(2.68)).
This prevents the momenta from being integrated out in the usual way, and a
configuration space sum over histories of the form \(1.18) is not obviously
obtained.

The relativistic particle is frequently studied as a toy model for quantum
gravity, and this is indeed part of the motivation for the study described in
this paper. In such investigations, it is often stated that the problem with
the Klein-Gordon equation is that the standard inner product is indefinite, and
thus it is necessary to discard half of the solutions [\cite{kuch}].
We would like to
point out, however, that it is not necessary to view the problem in this way.
As we have seen, there does in fact exist a positive definite inner product on
the set of all solutions to the Klein-Gordon equation, namely \(2.77). It is
therefore not necessary to discard any of the solutions if one uses this inner
product. Of course, the real problem with the Klein-Gordon equation is that
it is not possible to sort out the solutions into positive and negative
frequency, except in the simplest of situations. This problem is
present whatever view one takes.

\head {\bf 3. The Path Decomposition Expansion}
\taghead{3.}

Our ultimate task is to derive the various
relativistic composition laws from the sum
over histories \(2.50). The sum over histories for the relativistic particle
readily reduces to the proper time representation \(1.19). The derivation of
the
desired composition law is therefore intimately related to that of factoring a
sum over histories of the non-relativistic form \(1.11) across an arbitrary
surface in configuration space. As noted in the Introduction, the solution
to this problem already exists, and goes by the name of the path decomposition
expansion (PDX). In this section, we will describe this result, and give
a rigorous derivation of it.

\subhead {\bf 3(A). The PDX as a Partitioning of Paths}

Consider non-relativistic quantum mechanics in a configuration space
${\cal C}$ (here
taken to be $\R^n$), described by a propagator $ g( \x^{\pp}, T | \x', 0 )$.
The propagator may be expressed as a sum over histories, which we write,
$$
	g( \x^{\pp}, T | \x', 0 )
		= \int { \cal D} \x(t) \exp \left( i \int_0^T dt
		\left[ \half M \dot \x^2 - V(\x) \right] \right)\ .
	\eqno(3.1)
$$
The sum is taken over all paths in configuration space, $\x(t)$,
satisfying the boundary conditions $\x(0) = \x'$ and $\x(T) = \x^{\pp}$.
Denote this set of paths by $p(\x',0 \ria \x^{\pp},T) $.

Let $\Sigma$ be a surface between $ \x^{\pp} $ and $\x'$. It
therefore divides ${\cal C}$ into two parts, $\C_1$ and $\C_2$, say, with
$ \x' \in \C_1 $ and $ \x^{\pp} \in \C_2 $. $\Sigma$ may be closed or infinite.
We would like to factor the sum-over-histories across the surface $\Sigma$.

Consider the set of paths $p(\x',0 \ria \x^{\pp},T) $. Every path
crosses $\Sigma$ at least once, but will generally cross it many times
(see Fig. 3).
Unlike surfaces of constant time in spacetime, therefore, the position
of crossing does not label each path in a unique and unambiguous manner.
However, each path is uniquely labeled by the time and location
of its {\it first} crossing of $ \Sigma $. This means that there exists a
partition of the paths according to their time $t$ and location $\x_{\sigma}$
of first crossing (see Fig. 4). We write,
$$
	p(\x',0 \ria \x^{\pp},T) = \bigcup_{\x_{\sigma} \in \Sigma}
		\ \ \bigcup_{t \in [0,T]} \
		p(\x', 0 \ria \x_{\sigma}, t \ria \x^{\pp}, T )
%	\eqno(3.2)
$$
and
$$
	p(\x', 0 \ria \x_{\sigma}, t \ria \x^{\pp}, T ) \cap
		p(\x', 0 \ria \y_{\sigma}, s \ria \x^{\pp}, T ) = \emptyset,
		\quad {\rm if} \quad \x_{\s} \ne y_{\s}, \quad t \ne s\ .
%	\eqno(3.3)
$$
Each path in each part $ p(\x', 0 \ria \x_{\sigma}, t \ria \x^{\pp}, T ) $ of
the partition may then be split into two pieces:

\item{(i)} a restricted path lying entirely in $\C_1$, beginning at $\x'$ at
time $0$ and ending on $\Sigma $ at $ \x_{\s} $ at its first-crossing time $t$;

\item{(ii)} an unrestricted path exploring $\C_1$ and $\C_2$, beginning on
$\Sigma$ at $\x_{\s}$ at time $t$ and ending at $\x^{\pp}$ at time $T$.

This suggests that there exists a
composition of \(3.1) across $\Sigma$, consisting of a restricted propagator
in $\C_1$ from $(\x', 0)$ to $(\x_{\s}, t)$, composed with a standard
unrestricted propagator in $\C$ from $(\x_{\s}, t)$ to $(\x^{\pp},T)$,
with summations over both $ \x_{\s} $ and $ t $. There is indeed such
a composition law. It is the path decomposition expansion
[\cite{AK},\cite{vB}]:
$$
	g(\x^{\pp}, T| \x', 0 ) = \int_0^T dt \int_{\Sigma} d\s
		\left.\ g(\x^{\pp}, T| \x_{\s} , t )
		\ {i \over 2M} \ {\bf n} \cdot {\bf \nabla}
		g^{(r)} (\x,t| \x', 0 ) \right\vert_{\x=\x_{\s}}\ .
	\eqno(3.4)
$$
Here, $d \s$ is the integration over the surface $\Sigma$. The quantity
$g^{(r)}$ is the restricted propagator in $\C_1$, and satisfies the boundary
condition that it vanish on $\Sigma$. Its normal derivative
$ {\bf n} \cdot {\bf \nabla} g^{(r)} $, however, does not vanish on $\Sigma$.
Also note that ${\bf n}$ is defined to be
the normal to $\Sigma $ pointing {\it away} from
the region of restricted propagation, in this case $\C_1$. The reason for the
appearance of the normal derivative term will become fully apparent in the
rigorous derivation given below. For the moment we comment that it is related
to the fact that we are interested in restricted propagation to a final point
which actually lies on the boundary.

The path decomposition expansion is central to this paper, and we will be
making heavy use of it in what follows.

We now record some useful closely related results. First of all,
it is also possible to partition the paths according to their {\it last}
crossing times. This would lead to the composition law,
$$
	g(\x^{\pp}, T| \x', 0 ) = - \ \int_0^T dt \int_{\Sigma} d\s
		\ {i \over 2M} \ {\bf n} \cdot
		{\bf \nabla} \ g^{(r)} (\x^{\pp}, T| \x , t )
		\Big\vert_{\x=\x_{\s}}\ g (\x_{\s} ,t| \x', 0 )
	\eqno(3.5)
$$
where $t$ is the last crossing time. The overall minus sign arises because the
restricted propagator is now in the region $\C_2$, and the normal ${\bf n}$
(whose definition is unchanged) now points {\it into} the region of restricted
propagation.

Secondly, it is of interest to consider the case in which the surface
$\Sigma $ does not lie between the initial and final points,
$\x', \x^{\pp} \in \C_1 $, say. Then it is no longer true that every
path crosses $\Sigma$. In this case, one first partitions the paths
into paths that never cross $\Sigma$ and paths that always cross. The
paths that always cross may then be further partitioned as above. The
sum over paths which never cross simply yields a restricted propagator
in the region $\C_1$ that vanishes on $\Sigma$. One thus obtains,
$$
\eqalignno{
	g(\x^{\pp}, T| \x', 0 ) =& g^{(r)} (\x^{\pp}, T| \x', 0 )
	\cr
	+ & \left.\int_0^T dt \int_{\Sigma} d\s
		\ g(\x^{\pp}, T| \x_{\s} , t ) \
		{i \over 2M} \ {\bf n} \cdot {\bf \nabla}
		g^{(r)} (\x,t| \x', 0 ) \right\vert_{\x=\x_{\s}}
	&(3.6)
	\cr
}
$$
where $t$ is the first crossing time, and $g^{(r)}$ is the restricted
propagator in $\C_1$. ${\bf n}$ is again the normal pointing away from
$\C_1$. Similarly, in the case that the paths are partitioned according
to their final crossing times, one obtains
$$
\eqalignno{
	g(\x^{\pp}, T| \x', 0 ) =& g^{(r)} (\x^{\pp}, T| \x', 0 )
	\cr
	+ & \int_0^T dt \int_{\Sigma} d\s
		\left. \ {i \over 2M} \ {\bf n} \cdot {\bf \nabla} \
		g^{(r)} (\x^{\pp}, T| \x , t )
		\right\vert_{\x=\x_{\s}}\ g (\x_{\s} ,t| \x', 0 ) \ .
&(3.7) \cr}
$$
Here $t$ is the final crossing time, $g^{(r)}$ is again the restricted
propagator in $\C_1$ and ${\bf n}$ is again the normal pointing away
from $\C_1$. Note that there is no minus sign in the second term in Eq.
\(3.7),
in contrast to Eq. \(3.5). This is because in both \(3.6) and \(3.7),
the region of restricted propagation is $\C_1$ in each case, whereas in
\(3.4) and \(3.5), it is $\C_1$ and $\C_2$, respectively. These subtle
differences will turn out to be significant in Section 4.

\def\dt{\Delta \tau}
\def\dx{\Delta x}
\def\e{{\bf e}}

\subhead {\bf 3(B). A Lattice Derivation of the PDX}

The sum over histories \(3.1) must be regarded as no more than a formal
expression. Certain formal properties can sometimes be deduced from \(3.1)
as it stands, but care is generally necessary. In particular, the
path decomposition expansion cannot be derived directly from the sum over
histories without recourse to a more precise mathematical definition. The
purpose of this section, therefore, is to give a rigorous derivation of the
PDX from a properly-defined sum over histories.

Real time path integrals cannot be rigorously defined [\cite{C}], so we
first rotate to the imaginary time (Euclidean) version, by writing
$t= -i \tau $ (note that the Euclidean time $\tau$ bears no relation to the
parameter time $\tau$ of the previous section), yielding
$$
	g_E (\x^{\pp}, \tau | \x', 0 )
		= \int {\cal D} \x (\tau') \ \exp \left(
		- \int_0^{\tau} d \tau' \ \left[ \half
		M\dot \x^2 + V(\x) \right] \right)\ .
	\eqno(X.1)
$$
Euclidean sums over histories may be rigorously defined as the
continuum limit of a discrete sum over histories on a spacetime lattice. The
discrete sum over histories is then viewed as a sum of probability measures on
the space of paths on the lattice for some suitable stochastic process.
To illustrate the key features of the derivation of the PDX, we will
first consider the case of the free particle, $V(\x)=0$, and define
the sum over histories using one particularly simple stochastic process,
namely the random walk.

Consider a spacetime lattice with temporal spacing $\dt $ and spatial spacing
$\dx$. We follow the methods of Itzykson and Drouffe [\cite{ID}]. Let the
$n$-dimensional spatial lattice be generated by $n$ orthonormal vectors,
$\e_{\mu}$, with $\e_{\mu} \cdot \e_{\nu} = \dx^2 \ \delta_{\mu\nu} $. Each
site
is located at $\x = x^{\mu} \e_{\mu}$, where the $x^{\mu}$ are integers.

We propose to regard $ g_E (\x^{\pp}, \tau | \x', 0 ) $ as the continuum limit
of a probability density $p(\x^{\pp},\tau|\x',0)$. This quantity is defined to
be the probability density that in a random walk on the spacetime
lattice, the system (a particle, say) will be found at $\x^{\pp}$ at time
$\tau$ given that it was at $\x'$ initially. On the lattice it is meaningful to
talk about the probability of an individual history from $\x'$ at time zero to
$\x^{\pp}$ at time $\tau$. The probability density $p(\x^{\pp},\tau|\x',0)$ is
therefore given by the sum of the probabilities for the individual histories
connecting the initial and final points. Formally we write,
$$
	p(\x^{\pp},\tau|\x',0) = \sum_{histories} \ p (history)\ .
	\eqno(X.3)
$$
It is in this sense that it corresponds to a sum over histories.

$p(\x^{\pp},\tau|\x',0)$ satisfies the relations,
$$
\eqalignno{
	(\dx)^n \ p(\x^{\pp}, 0 | \x', 0 ) &=  \delta_{\x^{\pp},\x'}
	&(X.4)
	\cr
	(\dx)^n \ \sum_{\x^{\pp} }  \ p(\x^{\pp}, \tau | \x', 0 ) &= 1
	&(X.5)
	\cr
}
$$
where $\delta_{\x^{\pp},\x'}$ denotes a product of Kronecker deltas.
The factors of $(\dx)^n$ enter because $p$ is a density. Eq.
\(X.4) expresses the
initial condition, and \(X.5) says that the particle must be somewhere at time
$\tau$.

In a random walk, the probabilities of stepping from any one site to any one of
the adjacent sites are all equal, and equal to $1/2n$ in $n$ spatial
dimensions. The probability of an entire history in \(X.3) is then just $1/2n$
raised to the power of the number of steps in that history. Proceeding in this
way, one may evaluate \(X.3) and calculate the probability density $p$.
However,
we find it instead to be more convenient to calculate $p$ using the
recursion relation,
$$
\eqalignno{
	p(\x, \tau + \dt| \x',0) - p(\x,\tau|\x',0) =
		{ 1 \over 2n} \sum_{\mu=1}^n \ \left[
		\right. p(\x+\e_{\mu},\tau| & \x',0)
		+  p(\x-\e_{\mu},\tau|\x',0)
	\cr
	& - 2 \ p(\x,\tau|\x',0) \left. \right]\ .
	&(X.6)
	\cr
}
$$
This relation follows from the fact that if the walker is on site
$\x$ at time $\tau+\dt$, he must have been on one of the immediately
adjacent sites at time $\tau$. Eq. \(X.6) is a discrete version of
the diffusion equation. It may be solved by Fourier transform, yielding
the result
$$
	p(\x^{\pp},\tau|\x',0)
		= \int_{-\pi/\dx}^{\pi/\dx} \ {d^n {\bf k} \over
		(2 \pi)^n } \ e^{i {\bf k} \cdot
		(\x^{\pp} - \x' )} \ \left( {1 \over n}
		\sum_{\mu=1}^n \ \cos \dx k_{\mu} \right)^{\tau/\dt}\ .
%	\eqno(X.7)
$$
Taking the continuum limit, $\dt, \dx \ria 0 $, and holding fixed the
combination,
$$
	{ (\dx)^2 \over 2 n \dt } = { 1 \over 2M}
	\eqno(X.8)
$$
one obtains,
$$
	g_E (\x^{\pp},\tau|\x',0)
		= \left( {M\over 2 \pi \tau} \right)^{n \over 2}
		\exp \left( - {M(\x^{\pp} -\x')^2 \over 2 \tau } \right)
%	\eqno(X.9)
$$
where we use $g_E$ to denote the continuum limit of $p$.
The diffusion limit of this stochastic process therefore yields the
Euclidean propagator for the free non-relativistic particle of mass $m$.

Armed with a more precise notion of a discrete sum over histories, we may now
proceed to the derivation of the PDX. For simplicity, we first restrict
attention to the case in which the intermediate surface $\Sigma$ is
flat\note{The validity of the PDX for arbitrary surfaces has
been demonstrated in Ref. [\cite{vB}] by the use of a generalized Green's
theorem.}.
We view $p(\x^{\pp},\tau|\x',0)$ as a sum
of the probabilities for each path on the lattice from the initial to the final
point. As described in Section 3(A), the paths may be partitioned according to
their position $\x_{\s}$ and time $\tau_c$ of first crossing of an intermediate
surface $\Sigma$. We therefore expect a composition law on the lattice
expressing the statement, ``The probability of going from $\x'$ at time zero
to $\x^{\pp}$ at time $\tau$ is the sum over $\x_{\s}$ and $\tau_c$
of the probabilities of going from the initial point to final point crossing
the surface $\Sigma$ for the first time at time $\tau_c$ at
the point $\x_{\s}$''. The composition law is,
$$
	p(\x^{\pp},\tau|\x',0) =  (\dx)^n \ \sum_{\x_{\s} \in \Sigma}
		\ \sum_{\tau_c=0}^{\tau} \
		\tilde p(\x^{\pp},\tau| \x_{\s}, \tau_c )
		\ q(\x_{\s}, \tau_c |\x',0)\ .
	\eqno(X.10)
$$
Here, $q(\x_{\s}, \tau_c |\x',0)$ is defined to be a lattice sum over
paths which never cross $\Sigma$ but end on it at position $\x_{\s}$ at
time $\tau_c$. After reaching the surface at the point $\x_{\s}$ at time
$\tau_c$, the paths must then actually step across it, by definition of
the partition.
The quantity $ \tilde p(\x^{\pp},\tau| \x_{\s}, \tau_c )$ is therefore
a lattice sum over all paths from the surface to the final point, but
with the restriction that the very first step moves off the surface in
the normal direction. It is therefore given by,
$$
	\tilde p(\x^{\pp},\tau| \x_{\s}, \tau_c ) = {1 \over 2n}
		\ p(\x^{\pp},\tau| \x_{\s}+ \dx \ {\bf n}, \tau_c + \dt )
	\eqno(X.11)
$$
since $1/2n$ is the probability of stepping off the surface, and
$\ p(\x^{\pp},\tau| \x_{\s}+ \dx \ {\bf n}, \tau_c +\dt)$ is the probability
of going from the point just off the surface to the final point.
Strictly the sum over $\tau_c$ should not begin at zero, because on the
lattice it takes a finite amount of time for the first path to reach
the surface, but this time interval goes to zero in the continuum limit.

Because $q(\x^{\pp},\tau |\x',0)$ is a sum over paths that never cross
$\Sigma$ (but may touch it), it will satisfy the boundary condition
$$
	q(\x_{\s}+ \dx \ {\bf n}, \tau_c |\x',0) = 0
	\eqno(X.12)
$$
where ${\bf n}$ is the normal to the surface.
That is, the probability of making one step beyond $\Sigma$ is zero.
Now write
$$
	q(\x_{\s}, \tau_c |\x',0) = q(\x_{\s}+ \dx \ {\bf n}, \tau_c |\x',0)
		- \dx \left[ {q(\x_{\s}+ \dx \ {\bf n},
		\tau_c |\x',0) - q(\x_{\s}, \tau_c
		|\x',0) \over \dx} \right] \ .
	\eqno(X.13)
$$
The boundary condition \(X.12) implies that the first term vanishes. The
part of the second term in square brackets converges to
the normal derivative of $q$ in the continuum limit. Inserting
this in \(X.10), one obtains, with some rearrangement,
$$
\eqalignno{
	p(\x^{\pp},\tau|\x',0) = & \sum_{\x_{\s} \in \Sigma}
		\ (\dx)^{n-1} \ \sum_{\tau_c=0}^{\tau}
		\ \dt \ \ p(\x^{\pp},\tau| \x_{\s}+\dx
		\ {\bf n}, \tau_c +\dt )
	\cr
	& \times \ { (\dx)^2 \over 2n \dt}
		\ \left[ {q(\x_{\s}+ \dx \ {\bf n},
		\tau_c |\x',0) - q(\x_{\s}, \tau_c
		|\x',0) \over \dx} \right]\ .
	&(X.14)
	\cr
}
$$
Now, using the continuum limits
$$
	\sum_{\x_{\s} \in \Sigma } \ (\dx)^{n-1} \ria \int_{\Sigma}  d\s,
		\quad\quad\quad\quad
		\sum_{\tau_c=0}^{\tau} \ \dt \ria \int_0^{\tau} d \tau_c
	\eqno(X.15)
$$
and using \(X.8), we derive
$$
	g_E(\x^{\pp},\tau|\x',0)
		= \int_0^{\tau} d \tau_c \ \int_{\Sigma}  d\s
		\ g_E(\x^{\pp},\tau|\x_{\s}, \tau_c)
		\ {1 \over 2M} \ {\bf n}
		\cdot {\bf \nabla} g_E^{(r)}(\x, \tau_c| \x', 0 )
		\Big\vert_{\x=\x_{\s}}
	\eqno(X.16)
$$
This is the Euclidean version of the path decomposition expansion.
The desired result \(3.4) is then readily obtained by continuing back to
real time. The closely related results \(3.5)--\(3.7) are derived in a similar
manner.

It is perhaps worth noting that this result cannot not be derived from formal
manipulation of the continuum sum over histories \(3.1). Each part of the
composition law \(X.10) is well-defined and non-zero on the lattice, but not
every part has a continuum analogue. In particular $q(\x_\s,\t_c|\x,0)$,
where $\x_\s$ is on $\Sigma$, formally goes to zero
in the continuum limit. The desired result arises because the various parts of
\(X.10) fortuitously conspire to give a result which is well-defined in the
continuum limit, even though the separate parts may not be.

Now consider the case of non-zero potential, $V(\x) \ne 0 $.
We will argue that
the inclusion of a potential does not affect the key points of the derivation
of the path decomposition expansion. The random walk process described above
supplies a {\it measure} on the set of paths on the lattice. (In fact it is an
important result that it also defines a measure in the continuum limit, but we
prefer to work on the lattice.) Using this measure, one can compute the
average value of various functions of the histories of the system. In
particular, it is a standard result that the amplitude \(X.1) may be defined
as the average value of $\exp\left(- \int d\tau V(\x(\tau)) \right)$ in
this measure [\cite{Kac}].

A different way of doing essentially the same calculation is more convenient
for our purposes.
The amplitude \(X.1) may be calculated directly by
constructing a measure on the set of paths different to that given above,
which includes the effect of the potential. A weight $w(history)$ may be
defined for each history, and the density $w(\x'',\t|\x',0)$ is again
$$
	w(\x'',\t|\x',0)=\sum_{histories} w(history).
$$
Loosely speaking, $w$ is defined by weighting the probability $p$ of going
from one lattice point to the next by $\exp\left( - \dt V(\x) \right)$. $w$
is of course no longer a probability density, and does not define a
stochastic process. It obeys the recursion relation
$$
\eqalign{
	w(\x, \tau + \dt| \x',0) - w(\x,\t|,\x',0)=
		{ 1 \over 2n} \sum_{\mu=1}^n \ \left[
		w(\x+\e_{\mu},\tau|\x',0) \right.
	\cr
	+ \left. w(\x-\e_{\mu},\tau|\x',0)-2\ w(\x,\t|\x',0) \right]
		+\dt V(\x) w(\x,\t|\x',0)
	\cr
}
	\eqno(3.77)
$$
which differs from \(X.6) in that the `walker' may now stay at site $\x$
with a weight $\dt V(\x)$. This recursion relation yields the Euclidean
Schr\"odinger equation with potential $V(\x)$  in the continuum limit, as
expected.

The issue is now to determine whether the derivation, \(X.10)--\(X.16) goes
through for $w$ as it did for $p$. It is relatively easy to see that it
will. The quantities analogous to
$\tilde{p}$ and $q$ are defined in the obvious way,
and all the steps go through as before. The important point is that \(X.10)
and \(X.11) are not modified, since the weight for stepping off the surface
is still $1/2n$, as may be seen from the recursion relation \(3.77).

An equivalent approach is to rescale
the weights $w$ so that they describe a stochastic process, and can be
regarded as probability densities [\cite{Y}].
The random walk
is then characterised by a non-zero drift, that is by unequal probabilities
of stepping in different directions due to the asymmetry of the potential.
In the continuum limit, the rescaled
$w$ satisfies a Fokker-Planck equation. A composition law
involving an object analogous to $q$ may then be derived, which is a
rescaled version of the path decomposition expansion. We will not pursue
this here\note{It
is also possible to prove the path decomposition
expansion for non-zero potential by using the Wiener measure on the
set of all continuous Brownian paths [\cite{sam}] (for a description of the
Wiener measure, and its role in the sum-over-histories, see for example
Ref. \cite{SB}).}.

\subhead{\bf 3(C). An Important Simplification}

The restricted propagator appearing
in \(3.4)--\(3.7) is somewhat inconvenient and for our purposes
it is useful to re-express it in terms of the usual propagator [\cite{S}].
This is certainly possible if the potential $V(\x)$ in \(3.1) possesses a
translational symmetry in a direction that we shall refer to as $x^0$, and
$\Sigma$ is a surface of constant $x^0$.
To this end, consider $ g^{(r)} (\x,t| \x', 0 ) $ in \(3.4), where both
$\x$ and $\x'$ are in $\C_1$, the region of restricted propagation.
By the imposed
symmetry of the random walk, it is possible to rewrite the restricted
propagator as
$$
	g^{(r)} (\x'',t| \x', 0 )  =
		g (\x'' ,t| \x', 0 )
		- g (\x_{\s}+({\x_{\s}-\x''}) ,t| \x', 0 )
	\eqno(3.8)
$$
where $\x_\s$ is the point on $\Sigma$ closest to $\x''$.
This is of course just the familiar method of images. That this is
equivalent to a restricted sum over paths may be seen as follows. The full
propagator is given by a sum over all paths from initial to final point.
The sum over all paths $g$ may be written as  a sum over paths that never
cross the surface, $g^{(r)}$, plus a sum over paths that do cross the
surface at least once, $g^{(a)}$ ({\it c.f.} Eq. \(3.6)). The paths that
cross have a last crossing position. Because of the symmetry, the segment
of the path after the last crossing may be reflected about the surface
without changing the value of the sum over paths (see Fig. 5).
$g^{(a)}$ is therefore
equal to the sum over all paths from the initial point to the reflection
about the surface of the final point. Hence, with a little rearrangement,
one obtains \(3.8).

Given \(3.8),
the normal derivative of the restricted propagator on $\Sigma$, which is
the quantity that appears in \(3.4), is just
$$
	{\bf n} \left.\left.\cdot {\bf \nabla} g^{(r)} (\x,t| \x', 0 )
		\right\vert_{\x=\x_{\s}}
		= 2 \ {\bf n} \cdot {\bf \nabla} g (\x,t| \x', 0 )
		\right\vert_{\x=\x_{\s}}\ .
	\eqno(3.10)
$$
We conclude that in the special case of a symmetric potential and a
flat surface, \(3.4) becomes
$$
	g(\x^{\pp}, T| \x', 0 ) = \int_0^T dt \int_{\Sigma} d\s
		\left.\ g(\x^{\pp}, T| \x_{\s} , t )
		\ {i \over M} \ {\bf n} \cdot {\bf \nabla}
		g(\x,t| \x', 0 ) \right\vert_{\x=\x_{\s}}
	\eqno(3.9)
$$
and likewise for \(3.5).
We will use this result in all subsequent applications of the PDX.

The analysis so far is for flat surfaces $\Sigma$. \(3.10) will also follow
for curved surfaces (in flat configuration spaces with constant potential),
because the analysis leading to it is essentially local.

\head {\bf 4. Derivation of Relativistic Composition Laws}
\taghead{4.}

We now show how the path decomposition expansion is used to derive
the relativistic composition laws for certain Green functions.

\subhead {\bf 4(A). Composition Laws for $G_F$, $G^+$ and $G^-$}

Consider the Feynman Green function. As discussed earlier, its
sum-over-histories representation readily reduces to,
$$
	iG_F(x^{\pp}|x') = \int_0^{\infty} dT \ g(x^{\pp}, T| x', 0)\ .
	\eqno(4.1)
$$
Here $g(x^{\pp}, T| x', 0)$ is a propagator of the non-relativistic
type, and is given by a sum-over-histories of the form \(3.1), but
with $2M=1$, $V(x)=m^2$ (which means that the results of section 3(C)
apply for any flat surface), and with $\dot \x^2$ replaced by
$ \dot x^{\mu} \dot x^{\nu} \eta_{\mu\nu} $, where $\eta_{\mu\nu}$ is the
Minkowski metric, with signature $(+---)$. It therefore obeys the Schr\"odinger
equation,
$$
	\left(-i{\partial \over \partial T}+\Box_{x^{\pp}}+ m^2  \right)
		g(x^{\pp}, T| x', 0) = 0
%	\eqno(4.2)
$$
subject to the initial condition
$$
	g(x^{\pp}, 0| x', 0) = \delta^{(4)} ( x^{\pp} - x' )\ .
%	\eqno(4.3)
$$
{}From this, it readily follows that $G_F(x''|x')$ satisfies Eq. \(2.1).
An explicit expression for $ g(x^{\pp}, T| x', 0) $
is readily obtained:
$$
	g(x^{\pp},T|x',0)
		= {1\over (2\pi iT)^2}\exp \left(-i{(x^{\pp}-x')^2
		\over 4T}-im^2 T\right)\ .
%	\eqno(4.5)
$$

These basics out of the way, we may now derive the composition law. Consider
first the case in which the initial and final points are on opposite sides of
the surface $\Sigma$. Apply the path decomposition expansion \(3.4) to \(4.1).
One obtains,
$$
	G_F(x^{\pp}|x') =
		\ -i \ \int_0^{\infty} dT \ \int_0^T dt \ \int_{\Sigma}
		d \s \ g(x^{\pp}, T| x, t)
		\ 2i \ \ura{\partial_n} \ g( x, t | x', 0)\ .
	\eqno(4.6)
$$
Here, $ \ura{\partial_n} $ denotes the normal derivative pointing away
from $x'$ and operating to the right, and we have used \(3.10) to express
the derivative of the restricted propagator in terms of the unrestricted
propagator. Also, we use a simple $x$ to denote the coordinates in the
surface $\Sigma$. Now, in the integrals over time, one may perform
the change of coordinates $v = T-t$, and $u=t$. Eq. \(4.6) then becomes,
$$
	G_F(x^{\pp}|x') = \ 2 \ \int_0^{\infty} dv \ \int_0^{\infty} du
		\ \int_{\Sigma} d \s \ g(x^{\pp}, v | x, 0 )
		\ \ura{\partial_n} \ g( x, u | x', 0)\ .
	\eqno(4.7)
$$
Comparing with \(4.1), it is then readily seen that
$$
	G_F(x^{\pp}|x') = \ -2 \ \int_{\Sigma} d \s \ G_F(x^{\pp}|x)
		\ \ura{\partial_n} \ G_F(x|x') \ .
	\eqno(4.8)
$$
Although this is a correct property of the Feynman Green function, it
is not quite the expected result. Furthermore, it does not manifestly
exhibit the usual property of independence of the location of the
factoring surface. To this end, we repeat the above with \(3.5)
instead of \(3.4), obtaining,
$$
	G_F(x^{\pp}|x') = \ 2 \ \int_{\Sigma} d \s \ G_F(x^{\pp}|x)
		\ \ula{\partial_n} \ G_F(x|x') \ .
	\eqno(4.9)
$$
Finally, averaging \(4.8) and \(4.9) leads to the desired result:
$$
	G_F(x^{\pp}|x') = \ -\int_{\Sigma} d \s \ G_F(x^{\pp}|x)
		\ \ulra{\partial_n} \ G_F(x|x') \ .
	\eqno(4.10)
$$

Define $iG_F(x^{\pp},x')$ to be $G^+(x^{\pp},x')$ when $x^{\pp}$ is
in the future cone of $x'$, and to be $G^-(x^{\pp},x')$ when $x^{\pp}$ is
in the past cone of $x'$. Then it readily follows that $G^+$ and $G^-$ each
satisfy suitably modified versions of \(4.8) and \(4.9) and hence their
composition laws \(2.60).

Now consider the case in which the initial and final points lie on the
same side of the surface $\Sigma$. We therefore apply the path decomposition
expansions \(3.6), \(3.7). Direct application of either of these expressions
to Eq. \(4.1), does not lead to an obviously useful result, since
it still involves a restricted propagator. However, equating \(3.6) and
\(3.7), one obtains
$$
\int_0^T dt \ \int_{\Sigma} d \s \ g(x^{\pp}, T| x, t) \ \ura{\partial_n}
\ g( x, t | x', 0) =
\int_0^T dt \ \int_{\Sigma} d \s \ g(x^{\pp}, T| x, t) \ \ula{\partial_n}
\ g( x, t | x', 0)\ .
	\eqno(4.11)
$$
Suppose that $x^{\pp}$ is in the future cone of $x'$, which in turn lies
to the future of the surface $\Sigma$. Then performing the integral over
$T$ in \(4.11) leads to the result,
$$
\int_{\Sigma} d \s^\mu \ G^+ (x^{\pp}|x)
\ \ulra{\partial_\mu} \ G^- (x|x')  \ = \ 0
\eqno(4.12)
$$
demonstrating the expected orthogonality of $G^+$ and $G^-$.

It might appear that the above derivation of the composition law is valid
for {\it any} choice of factoring surface. This impression would be false:
the derivation holds only for spacelike surfaces. To see this, note
that the integral representation of the Feynman Green function
\(4.1) is properly defined only in the Euclidean regime. The Euclidean
version of \(4.1) is obtained by rotating both the parameter time $T$
and the physical time $x^0$. Write $T_E = iT $ and $x_E^0 = i x^0 $.
The first rotation is just a matter of distorting the integration contour
in \(4.1) and does not change the result of evaluating the integral. Indeed,
\(4.1) may be {\it defined} by an integral over real $T_E$. The second
rotation actually changes the answer, so needs to be rotated back
afterwards. Performing the rotations, one obtains,
$$
g_E (x^{\pp}, T_E | x', 0 ) = {1 \over (2 \pi T_E)^2 }
\exp \left( - {1 \over 4T_E} \left[ (x_E^0)^2 + {\bf x}^2 \right]
- m^2 T \right)
\eqno(4.66)
$$
for the time-dependent propagator, where $x$ denotes $x^{\pp} - x'$ for both
its time and space components. The Euclidean path decomposition expansion
\(X.16) is then clearly well-defined for \(4.66) -- the integral over the
surface $\Sigma$ is clearly convergent. A composition law for the Euclidean
Feynman propagator is therefore obtained across {\it any} surface. But
suppose now we try to continue back the Euclidean PDX \(X.16) to the
Lorentzian spacetime. Leave $T_E$ as it is, but continue back $x^0$. The
integrand, previously exponentially decaying in all directions, becomes
exponentially growing in the $x^0$ direction. This is not a problem if
the surface $\Sigma$ is spacelike, since $x^0$ is not integrated over.
It is a problem if $x^0$ is integrated over, which it would be if
$\Sigma$ is timelike. It follows that the Euclidean composition law, valid
for any surface, may be continued to a well-defined composition law for the
Lorentzian propagator only if the surface $\Sigma$ is spacelike in the
Lorentzian regime.

At this stage it is perhaps useful to summarize how we have arrived at the
results \(4.10), \(4.12) from the sum-over-histories. First the sum over
histories was written in the proper time representation, \(4.1). This is
essentially a partition of the set of all paths from $x'$ to $x^{\pp}$,
according to their total parameter time (which is effectively the same as their
length). Then the paths were further partitioned according to the parameter
time and position of their first (or last) crossing of $\Sigma$. The path
decomposition expansion then led to the desired result. In the final result
\(4.10), however, no reference is made to the parameter time involved in this
sequence of partitions; only the first (or last) crossing position $x$ is
referred to. In the results \(4.10), \(4.12), therefore, there is only one
partition of the paths that is important, namely the partition according to the
position $x$ of first or last crossing.
Differently put, suppose there existed a sum over histories representation of
$G_F$ referring only to the spacetime coordinates $x^{\mu}$, and not requiring
the explicit introduction of a parameter $t$. Then the composition law \(4.10)
could be derived by a single partitioning of the paths according to their first
or last crossing position.

By way of a short digression,
let us explore this idea further. Suppose one simply assumes that a sum over
histories representation of $G_F(x^{\pp}|x')$ is available, in which
there is a sum over all paths in spacetime from $x'$ to $x^{\pp}$. As
described above, one can therefore partition the paths according to their
first crossing position $x$ of an intermediate surface $\Sigma$. It is
therefore reasonable to {\it postulate} a relation of the form
$$
G_F(x^{\pp}|x') = \ \int_{\Sigma} d \s \ G_F(x^{\pp}|x)
\Delta(x|x')
%\eqno(4.13)
$$
where $ \Delta(x|x') $ is defined by a restricted sum over paths beginning
at $x'$ which never cross $\Sigma$, but end on it at $x$. Comparing
with Eq. \(4.8), or by explicit calculation, one has,
$$
\Delta(x|x') = 2 \ {\partial_n} G_F (x|x')\ .
\eqno(4.14)
$$
This gives rather intriguing representation of $ {\partial_n} G_F (x|x') $
in terms of a restricted sum over paths in spacetime.

It is also interesting to note that when $x^0>{x^0}^\p$,
by explicit calculation,
$$
2i{\partial_n} G_F (x|x') = G_{NW}(x|x')
%\eqno(4.15)
$$
where $G_{NW}$ is the Newton--Wigner propagator [\cite{H}].
Via \(4.14), this therefore
gives a novel path integral representation of the Newton--Wigner propagator. It
is novel because $G_{NW}$ is really a propagator of the Schr\"odinger type, and
is therefore normally obtained by a sum over paths moving
forwards in time, as we saw in section 2(B).
By contrast, in the path integral representation of
$\Delta(x|x')$, the paths move backwards and forwards in time, although
are restricted to lie on one side of the surface $\Sigma$ in which $x$
lies.

Note that using this representation of the Newton--Wigner propagator,
its composition law \(2.70) is easily derived. The sum over paths from $x'$
to $x''$ ending on $\Sigma$ and remaining below it,
may be partitioned across an
intermediate surface $\Sigma'$ according to the point $x_{\s'}$ of first
crossing of $\Sigma'$. That is
$$
\eqalign{
	& p(x'\to x'')=\bigcup_{x_{\s'}}p(x'\to x_{\s'}\to x'')
	\cr
	& p(x'\to x_{\s'}\to x'')\ \cap\ p(x'\to y_{\s'}\to x'')=
		\emptyset\qquad{\rm if}\qquad x_{\s'}\ne y_{\s'}\ .
	\cr
}
$$
The sum over paths factorises into a sum over paths from $x'$ to $x_{\s'}$,
ending on $\Sigma'$ and remaining below it, and over paths from $x_{\s'}$
to $x''$, ending on $\Sigma$ and remaining below it. This is precisely a
composition of type \(1.17), and leads directly to \(2.70).

These observations may merit further investigation. They are, however,
only incidental to the rest of this paper.

\subhead {\bf 4(B). Other Green Functions}

By integrating $T$ over an infinite range in \(1.19) the Green function
$G^{(1)}(x^{\pp}|x')$ is obtained. Let us therefore repeat the steps \(4.6)
to \(4.10) for this case. The integration over $T$ and $t$ in \(4.6) is now
$$
	\int_{-\infty}^{\infty} dT \ \int_0^T dt  =
		\int_0^{\infty} dT \ \int_0^T dt
		+ \int_{-\infty}^0 dT \ \int_0^T dt  \ .
	\eqno(4.16)
$$
The first term in \(4.16) leads to a composition of two Feynman Green
functions, as before. The second term can be cast in a similar form
by letting $T \ria - T$ and $t \ria -t$, which introduces an overall
minus sign, and using the fact that $g(x,-t|x',0)=g^*(x,t|x'0)$. One
thus obtains
$$
\eqalign{
	G^{(1)}(x^{\pp}|x')& = \ -i\int_{\Sigma} d \s \
		\left[ G_F(x^{\pp}|x)
		\ \ulra{\partial_n} \ G_F(x|x') - G_F^*(x^{\pp}|x)
		\ \ulra{\partial_n} \ G_F^*(x|x')\right]
	\cr
	&= \ i \int_{\Sigma} d \s^\mu \ \left[G^+(x^{\pp}|x)
		\ \ulra{\partial_\mu} \ G^+(x|x') - G^-(x^{\pp}|x)
		\ \ulra{\partial_\mu} \ G^-(x|x')\right]
	\cr
}
	\eqno(4.17)
$$
where $d\s^\mu$ and $\partial_n$ are defined as in section 2.
The result \(4.17) may seem somewhat trivial, since it follows from
\(4.10) and the use of
$$
\eqalign{
	G^{(1)}(x^{\pp}|x')
		&= i\left[G_F(x^{\pp}|x') - G_F^* (x^{\pp}|x')\right]
	\cr
	&= G^+(x^{\pp}|x')+G^-(x^{\pp}|x')\ .
	\cr
}
$$
However, the key point is that the
composition law \(2.80) for $G^{(1)}$ arises directly in the sum over
histories. The splitting into positive and negative frequency parts,
in the language of section 2, arises naturally from the identity \(4.16).

Finally, consider the causal Green function. In terms of $G^{\pm}$
it is defined by
$$
	iG(x^{\pp}|x')=G^+(x^{\pp}|x') -G^- (x^{\pp}|x')\ .
%	\eqno(4.19)
$$
Then it straightforwardly follows that $G$ obeys \(1.10), since $G^{\pm}$ obey
\(2.60) and
\(4.12). However, there is no natural, quantum mechanical derivation of the
composition law for $G$ directly from a sum over histories. This is because we
do not have a direct path-integral representation of $G$ -- only an indirect
one in terms of the path-integral representations of $G^{\pm}$ (which may be
read off from \(4.1)). The question of finding a direct sum-over-histories
representation of the causal propagator is, to the best of our knowledge, a
question for which no entirely satisfactory answer exists at present.
Indeed, as we conjectured in Section II, such a representation may not
exist.

\subhead {\bf 4(C). Why the Naive Composition Law Fails}

In the context of quantum gravity, and parameterized theories
generally, composition laws different in form to \(1.10) have occasionally
been proposed. In particular, a composition law of the form
$$
\G(x^{\pp}|x') = \int d^4 x \ \G(x^{\pp}|x) \ \G(x|x')
\eqno(4.20)
$$
has often been
considered [\cite{??}]. However, it is readily seen that there are
difficulties associated with \(4.20) [\cite{JJH2}].
The methods of this paper help to understand the reason why
it cannot hold as it stands.

Let us first illustrate the problem with \(4.20). Consider the proper time
representation \(1.19). It is a property of $g$ that,
$$
g(x^{\pp}, T^{\pp} + T' | x', 0 ) = \int d^4 x \ g (x^{\pp}, T^{\pp} | x, 0 )
\ g (x, T' | x', 0 )\ .
%\eqno(4.21)
$$
Integrating both sides over $T^{\pp}$ and $T'$, one obtains
$$
- \half \ \int du \int dv \ g ( x^{\pp}, u | x', 0 )
= \int d^4 x \ g ( x^{\pp} | x ) \ g ( x | x' )
\eqno(4.22)
$$
where we have introduced $ u = T^{\pp} + T' $, $v = T^{\pp} - T'$.
If $T$ is taken to have an infinite range, then $u$ and $v$ have an infinite
range, and the left hand side of
\(4.22) is equal to $G^{(1)}(x^{\pp}|x')$ multiplied by an infinite
factor. If $T$ is taken to have a half-infinite range, then things are yet
more problematic. In that case $v$ ranges
from $-u$ to $+u$, and the left-hand side of \(4.22) becomes,
$$
- \int_0^{\infty} du \ u \  g ( x^{\pp}, u | x', 0 )\ .
\eqno(4.23)
$$
This may converge, but it does not converge to the left-hand side of
\(4.20).

It should be clear from the discussion given in the Introduction that
\(4.20) should not be expected to hold. The reason is, quite
simply, that it does not correspond to a proper partitioning of the paths in
the sum-over-histories \(1.18). For in proposing an expression of the form
\(4.20), one is evidently contemplating a partitioning of the paths
$p(x' \ria x^{\pp})$ in which the paths are labeled according to an
intermediate spacetime point $x$ through which they pass. That is, the set of
all paths is regarded as the union over all $x$ of paths passing through $x$,
$$
p(x' \ria x^{\pp} ) = \bigcup_x p ( x' \ria x \ria x^{\pp} )\ .
%\eqno(4.24)
$$
But this is not a proper partition because it is not exclusive,
$$
p ( x' \ria x \ria x^{\pp} ) \cap ( x' \ria y \ria x^{\pp} ) \ne \emptyset,
\quad {\rm for} \quad x \ne y\ .
\eqno(4.25)
$$
It is not exclusive because passing through an intermediate point $x$ does not
prohibit the path from also passing through a different intermediate point
$y$. The intermediate spacetime point $x$ therefore does not supply the
paths with a unique and unambiguous label.

Of course, the exhaustivity condition \(4.25) is still in some sense true, but
the failure of the exclusivity condition means that there is a vast amount of
overcounting. It is this that leads to the divergent factor appearing
in \(4.23) in the case where $T$ takes an infinite range.

The fact that \(4.22) is equal to $G^{(1)}$ times an infinite factor is,
however,
suggestive. A similar feature was found in the Dirac quantisation of the
relativistic particle by Henneaux and Teitelboim [\cite{HT}]. They found
that for functions $\psi(x)=\bkb{x}{\psi}$ solving the Klein-Gordon
equation,
$$
	\bkb{\phi}{\psi}=\int d^4 x\ \phi^\dagger(x)\psi(x)
	\eqno(4.88)
$$
is a positive definite inner product independent of $x^0$, and with all the
necessary symmetry properties for an inner product on physical states. The
only problem with \(4.88) is that it is formally divergent. In fact, it is
equal to the inner product \(2.77) times a factor $\delta(0)$, which may be
removed in a Lorentz invariant way [\cite{HT,ct}]. This inner product may
therefore be of some value, despite the fact that it is not associated with
a partition of the sum over histories. It is yet to be seen whether these
features continue to hold in more complicated parameterized systems, such
as quantum gravity.

It is also clear that
a composition law in which the $d^4x$ in \(4.20) is replaced
by a $d^3x$ cannot be correct.
This would at first sight be
more in keeping with conventional quantum mechanics, since
one of the four $x^{\mu}$'s is time, and the composition law \(1.4) is at a
fixed moment of time. However,
it corresponds to contemplating a partition in which
the paths are labeled according to the position $x^i$ at which they cross a
surface $x^0=constant$. This fails because, as discussed
in the Introduction, it is not a proper partition. The paths typically cross
such a surface many times, and the crossing location does not label the paths
in a unique and unambiguous way.

We have seen in this paper that there is a partition that
does work, and does lead to the desired composition law. It is to
partition the paths according to their position of {\it first} crossing
of an intermediate surface.

\head {\bf 5. Discussion}
\taghead{5.}

The principal
technical aim of this paper was to show that the composition laws of
relativistic quantum mechanics may be derived directly from a sum over
histories by partitioning the paths according to their first crossing position
of an intermediate surface. We also derived canonical representations of the
propagators. These representations showed why the Hadamard Green function
$G^{(1)}$, which is the propagator picked out by the sum over histories, does
not obey a standard composition law. They also indicate why it is not obviously
possible to construct a sum-over-histories representation of the causal Green
function.

The notion of a sum over histories is extremely general. Indeed, as discussed
in the Introduction, it has been suggested that sum-over-histories formulations
of quantum theory are more general than canonical formulations. Central to
such generalized formulations of quantum mechanics is the notion of
a partition of paths. This simple but powerful notion replaces and generalizes
the notion of a complete set of states at a fixed moment of time used in
canonical formulations [\cite{harnew}].

In this paper we have investigated a particular aspect of the
correspondence between these two different approaches to quantum theory.
Namely, we demonstrated the emergence of the composition law from the
sum-over-histories approach, in the context of relativistic quantum mechanics
in Minkowski space. Quite generally, such a derivation will be an important
step in the route from a sum-over-histories formulation to a canonical
formulation in a reparameterization invariant theory.
We have admittedly not determined the exact status of the
composition law along this route. In particular, it is not clear
whether the existence of the composition law alone is a {\it
sufficient} condition for the recovery of a canonical formulation.
This would be an interesting question to pursue, perhaps taking as a
starting point the comments at the end of Section 2(C), on the
recovery of the canonical inner product given the propagator.
However, as
argued in the Introduction, it is at least a {\it necessary}
condition. It is therefore of interest to find a
situation in which this necessary condition is not satisfied.

Such a
situation is provided by the case of relativistic quantum mechanics
in curved spacetime backgrounds with a spacetime dependent mass term
({\it i.e.} a potential). Let us consider the generalization of our
results to this case.

The path decomposition expansion
\(3.4) is a purely kinematical result. As we have shown, it arises solely
from partitioning the paths in the sum over histories, and does not depend
on the detailed dynamics. We would therefore expect it to hold in a very
general class of configuration spaces, including curved ones. It follows
that for the relativistic particle, one would {\it always} expect a
composition law of the form
$$
\G(x^{\pp}|x')  = \int d \sigma^{\mu} \ \G(x^{\pp}|x) \ \partial_{\mu}
\ \G^{(r)}(x|x')
\eqno(5.1)
$$
where $\G^{(r)}$ is the restricted relativistic
propagator \note{One might reasonably ask which Green function is
involved in \(5.1), and to what extent it is defined in general
spacetimes. One can think of ${\cal G}$ as being the Feynman or
Hadamard propagators. These may be formally defined in {\it any}
spacetime, using the proper time representation, \(4.1), although
their interpretation in terms of positive and negative frequencies
is generally not possible [\cite{ct}].}.
In the case
of flat backgrounds, with constant potential, it was possible to express
the restricted Green functions in terms of unrestricted ones, using
\(3.8)-\(3.10). The important point, however, is that in general backgrounds,
and with arbitrary potentials, the steps \(3.8)-\(3.10) are not possible, and
a composition law of the desired type \(1.4) is not recovered. Of course,
\(5.1) is still a composition law of sorts, but $\G$ and $\G^{(r)}$ are
quite different types of object, and \(5.1) is not compatible with regarding
$\G(x^{\pp}|x')$ as a canonical expression of the form $\langle
x^{\pp}|x' \rangle $, since there is no known canonical representation for
$\G^{(r)}(x''|x')$.

What is needed for the steps \(3.8)-\(3.10) to work? The main issue
is understanding how the method of images can be generalized. First
of all, consider the case of one dimension with a potential. The
method of images yields the restricted propagator for any potential
which is symmetric about the factoring surface (actually a point in
one dimension). For example, the restricted propagator in $x>0$ for
the harmonic oscillator is readily obtained in this way. However, we
would like to obtain the restricted propagator on one side of {\it
any} factoring surface. The only potential invariant under
reflections about any point is a constant. So in one dimension,
\(3.10) follows if the potential is constant. Similarly, it is easy
to see that in flat spaces of arbitrary dimension, with a flat
factoring surface, \(3.10) will follow if the potential
is constant in the direction normal to the surface.

Now consider the case of curved spacetimes with a Lorentzian signature
(although our conclusions will not be restricted to this situation). From the
above, we have seen that the method of images will work if the
propagator is symmetric about each member of a family of factoring
surfaces. We will now argue that this will be true if there is a
timelike Killing vector.

Consider first the case of static spacetimes. This means there is a timelike
Killing vector field normal to a family of spacelike hypersurfaces.
It is therefore possible to introduce coordinates such that
$$
	g_{\mu\nu} dx^{\mu} dx^{\nu} = g_{00}(x^i)(dx^0)^2
		+ g_{kl}(x^i) dx^k dx^l
	\eqno(5.2)
$$
where $i,k,l=1,2,3$. The action in the sum-over-histories representation
of $g(x^{\pp}, T|x',0)$ is
$$
	S= \int_0^T \ dt \
		\left[ g_{\mu\nu} \dot x^{\mu} \dot x^{\nu} - V(x^{\mu})
		\right]
	\eqno(5.3)
$$
If the metric is of the form \(5.2) and if the potential is independent
of $x^0$, then the action \(5.3) will be invariant under reflections about
any surface of constant $x^0 $. It is reasonable to expect that the
path-integral measure will be similarly invariant, and hence the method of
images may be used to construct the restricted propagator in a region
bounded by $x^0 = constant$. We therefore expect \(3.10) to hold in static
spacetimes in which the potential is invariant along the flow of the
Killing field. We anticipate that this argument may be generalized to
stationary spacetimes (for which there is a Killing field that is not
hypersurface orthogonal), but we have not proved this.

What we find, therefore, is that the existence of a timelike Killing
vector field, along which the potential is constant, is a {\it
sufficient condition} for the existence of a composition law for
the sum over histories. We cannot conclude from the above argument
that it is also a necessary condition, although this is plausibly
true for a general class of configuration spaces, with the possible
exception of a limited number of cases in which special properties
of the space avoid the need for a Killing vector\note{Note that none
of these claims are in contradiction with the fact that the causal
Green function is well-defined and obeys a composition law on any
globally hyperbolic spacetime, even those possessing no Killing
vectors [\cite{fulling}]. The causal Green function does not appear
to have a sum-over-histories representation, whilst the above
conclusions specifically concern propagators generated by sums over
histories.}.  Modulo these possible exceptions, we have therefore
achieved our desired aim: we have found a situation -- spacetimes
with no Killing vectors -- in which the necessary condition for the
recovery of a canonical formulation from a sum over histories is
generally not satisfied.

This is a desirable conclusion: the existence of a timelike Killing
vector field is the sufficient condition for a consistent
one-particle quantization in the canonical theory (see Section 9 of
Ref. [\cite{kuch}] and references therein). Again, it is not
obviously a necessary condition because there could be spacetimes
with no Killing vectors but some special properties permitting
quantization in them.  We therefore find close agreement (although
not an exact correspondence) between our approach, in which the
canonical formulation is regarded as derived from a sum over
histories, and standard lore, in which it is constructed directly.

Turn now to quantum cosmology. As noted in the Introduction,
relativistic quantum mechanics is frequently used as a model for
quantum cosmology. In quantum cosmology, the wave function for the
system -- the universe -- obeys the Wheeler-DeWitt equation. This is
a functional differential equation which has the form of a
Klein-Gordon equation in which the four $x^{\alpha}$'s are replaced
by the three-metric field, $h_{ij}({\bf x})$, the ``mass'' term is
dependent on the three-metric, and the ``background'' (superspace,
the space of three-metrics) is curved.

As outlined in the Introduction, one may construct the propagator
between three-metrics. The object obtained is most closely analogous
to either the Feynman or the Hadamard propagators, as noted above.
One can then ask whether it obeys a composition law. An important
result due to Kucha\v{r} [\cite{kuch}] is that there are no Killing
vectors associated with the Wheeler-DeWitt equation. We therefore
find that there is no composition law for the propagator between
three-metrics generated by a sum over histories\note{This conclusion
does not exclude the possibility of the existence of a propagator
not given by a sum over histories which obeys a composition law, in
analogy with the causal Green function. Note also that the
propagation amplitude generated by a sum over histories will obey
the Wheeler-DeWitt equation [\cite{hh}], but this does
not imply the existence of a composition law or a canonical
formulation.}. It follows that we do not expect to recover a
canonical formulation. Again this is in agreement with standard lore
on the canonical quantization of quantum cosmology, which holds that
there is no consistent ``one-universe" quantization [\cite{kuch}].

Our final conclusions on the existence of a canonical scheme for
quantum cosmology are therefore not new. However, what has not been
previously appreciated, as far as we are aware, is the close
connection of this question with the question of the existence of a
composition law for the sum over histories.

Finally, we may comment on the suggestion of Hartle discussed in the
Introduction -- that the sum over histories is more general than the
canonical scheme. Our results are not inconsistent with this claim:
the absence of Killing vectors associated with the Wheeler-DeWitt
equation probably rules out a canonical quantization, but does
not obviously prevent the construction of sums over histories.
Of course, there still remains the question of how the sums over
histories are to be used to construct probabilities, {\it i.e.}, the
question of {\it interpretation}. This is a difficult question and
will not be addressed here.

We emphasize that these arguments are intended to be suggestive, rather
than rigorous. These issues will be considered in greater detail in future
publications.

\head {\bf Acknowledgements}

We would like to thank Arlen Anderson, Franz Embacher, Eddie Farhi,
Jeffrey Goldstone, Sam Gutmann, Jim Hartle, Chris Isham, Samir
Mathur and Claudio Teitelboim for useful conversations. We would
particularly like to thank Larry Schulman for useful conversations,
and for introducing us to the path decomposition expansion. J.J.H.
was supported in part by a Royal Society Fellowship. M.O. was
supported by an SERC (UK) fellowship while part of this work was
carried out.

\endpage

\references

\def\pr{{\sl Phys. Rev.\ }}

\def\jmp{{\sl J. Math. Phys.\ }}

\def\np{{\sl Nucl. Phys.\ }}
\def\pl{{\sl Phys. Lett.\ }}
\def\annp{{\sl Ann. Phys. (N.Y.)\ }}

\refis{ish} C. Isham, `Canonical quantum gravity and the problem of
time', Imperial College preprint TP/91-92/25, gr-qc/9210011.

\refis{hh} J. J. Halliwell and J. B. Hartle, \pr {\bf D43}, 1170 (1991).

\refis{AK} A. Auerbach and S.Kivelson, \np {\bf B257}, 799 (1985).
% The path decomposition expansion and multidimensional tunneling

%\refis{WH} U. Weiss and W. Haeffner, \pr {\bf D 27}, 2916 (1983).
% Complex-time path integrals beyond the stationary phase approximation:
% Decay of metastable states and quantum statistical metastability.
% Not cited

\refis{vB} P. van Baal, `Tunneling and the path decomposition expansion',
Utrecht Preprint THU-91/19 (1991).

\refis{G} J. Govaerts, {\sl Hamiltonian Quantisation and Constrained Dynamics},
(Leuven University Press, Leuven, Belgium, 1991).

%\refis{dw} B. De Witt, \pr {\bf 160}, 1113 (1967).

\refis{HT} M. Henneaux and C. Teitelboim, \annp {\bf 143}, 127 (1982).

\refis{HK} J. B. Hartle and K. V. Kucha\v{r}, \pr {\bf D 34}, 2323 (1986).

\refis{JJH} J.J. Halliwell, \pr {\bf D 38}, 2468 (1988).

\refis{JJH2} J. J. Halliwell, in {\sl Conceptual Problems of Quantum Gravity},
edited by  A. Ashtekhar and J. Stachel (Birkh\"auser, Boston, 1991).

\refis{S} L. Schulman and R. W. Ziolkowiski, in {\sl Path integrals from
meV to MeV}, edited by V. Sa-yakanit, W. Sritrakool, J. Berananda, M. C.
Gutzwiller, A. Inomata, S. Lundqvist, J. R. Klauder and L. S. Schulman
(World Scientific, Singapore, 1989).

\refis{Y} N. Yamada, {\sl Sci. Rep. T\^ohoku Uni., Series 8} {\bf 12}, 177
(1992).

\refis{J} H. Ikemori, \pr {\bf D 40}, 3512 (1989).

\refis{C} R. H. Cameron, \jmp {\bf 39}, 126 (1960).

\refis{H} J. B. Hartle, private communication.

\refis{sam} S. Gutmann, private communication.

\refis{SB} L. S. Schulman, {\sl Techniques and Applications of Path
Integration}
(Wiley, New York, 1982).

\refis{Schiff} L. I. Schiff, {\sl Quantum Mechanics} (McGraw--Hill, New York,
1955).

\refis{T} C. Teitelboim, \pr {\bf D 25}, 3159 (1982).

\refis{ID} C. Itzykson and J. M. Drouffe, {\sl Statistical Field Theory Vol. 1:
{}From Brownian Motion to Renormalization and Lattice Gauge Theory} (Cambridge
University Press, Cambridge, 1989).

\refis{??} The naive composition law, in the context of quantum gravity,
has appeared in numerous papers over the years. Just a few of them are,
S. W. Hawking, in {\sl General Relativity, An Einstein Centenary Survey},
eds S. W. Hawking and W. Israel (Cambridge University Press, Cambridge,
1979); J. A. Wheeler, in {\sl Relativity, Groups and Topology}, eds B. S.
DeWitt and C. M. DeWitt (Gordon and Breach, New York, 1963); S. coleman,
\np {\bf B310}, 643 (1988); S. Giddings and A. Strominger, \np {\bf B321},
481 (1988). Many of the above authors appreciated that it may be incorrect,
but not for the reasons given in this paper.
%Various authors: naive composition law

\refis{Kac} M. Kac, {\sl Probability and Related Topics in the Physical
Sciences} (Interscience, New York, 1959).

\refis{BFV} E. S. Fradkin and G. A. Vilkowiski, \pl {\bf 55B}, 224 (1975);
CERN Report No. TH-2332, 1977 (unpublished); I. A. Batalin and G. A.
Vilkowiski, \pl {\bf 69B}, 309 (1977).

\refis{kuch} K. Kucha\v{r}, `Time and interpretations of quantum gravity',
to appear in the {\sl Proceedings of the 4th Canadian conference on General
Relativity and Relativistic Astrophysics}, eds G. Kunstatter, D. Vincent
and J. Williams, World Scientific, Singapore, 1992.

\refis{fulling} S. Fulling, {\sl Aspects of Quantum Field Theory in Curved
Spacetime} (Cambridge University Press, Cambridge, 1989).

\refis{harnew} The idea that the sum over histories may supply more a general
formulation of quantum mechanics has been emphasized by Hartle. See for
example, J. B. Hartle, \pr {\bf D38}, 2985 (1988); \pr {\bf D44} 3173
(1991); in {\sl Quantum Cosmology and Baby Universes}, Proceedings of the
Jerusalem Winter School on Theoretical Physics, eds S. Coleman, J. Hartle,
T. Piran and S. Weinberg (World Scientific, Singapore, 1991); in
Proceedings of the {\sl International Symposium on Quantum Physics and the
Universe}, Waseda University, Tokyo, Japan (1992).

\refis{ct} C. Teitelboim, private communication.

\endreferences

\vfill\eject

\font\ten=cmr10 scaled 1100
\font\tw=cmr10 scaled 1200
\tolerance 10000

\def\strut{\hbox{\vrule height 14pt depth 6pt width 0pt}}
\def\superstrut{\hbox{\vrule height 25pt depth 20pt width 0pt}}
\baselineskip 12pt plus 1pt minus 1pt
\def\ll{\langle}
\def\rr{\rangle}
\def\one{{1\hskip -.30em1}}
\def\one{{\ten 1}\hskip -.30em{\tw 1}}
\def\lessgtr{\hbox{$ {     \lower.50ex\hbox{$>$}
                   \atop \raise.30ex\hbox{$<$}
                   }     $}  }
\vsize=7in
Table 1. The various Green functions, and their roles in
non-relativistic quantum mechanics. ${\cal G}\circ{\cal G}$
and ${\cal G}\times{\cal G}$ denote relativistic
and non-relativistic composition laws respectively. Unless otherwise
stated, sums over histories are over arbitrary paths in spacetime from
$x$ to $y$.
\vskip 2 cm

\hoffset=-.25in
$$
\hbox
{
\vbox
{
\offinterlineskip
\hrule\hrule
\halign
{
\strut
\vrule#\tabskip0.025in
&
	$#$\hfil
&\vrule#&
	$#$\hfil
&\vrule#&
	$#$\hfil
&\vrule#&
	$#$\hfil
&
\vrule#\tabskip 0.0in
\cr
&
	\omit\hfil\hbox{\bf Green F$^{\underline{\bf n}}$}\hfil
&&
	\omit\hfil\hbox{\bf Composition Law}\hfil
&&
	\omit\hfil\hbox{\bf Sum Over Histories}\hfil
&&
	\omit\hfil\hbox{\bf Canonical Rep.\ }$\ll x|y\rr$\hfil
&
\cr
\noalign{\hrule}
\superstrut
&
	\eqalign{& G^+(x,y) \cr & \cr & G^-(x,y)\cr}
&&
	\left. \eqalign{G^+ &= G^+\circ G^+ \cr &
	\cr G^-&= G^-\circ
	G^-\cr}\right\} G^+\circ G^-=0
&&
	\omit\hfil
	$\matrix{N>0 & x^0>y^0 \cr N<0 & x^0<y^0\cr
	&\cr N>0 & x^0<y^0\cr N<0 & x^0> y^0 \cr}$\hfil
&&
	\left. \eqalign{& p_0>0 \cr & \cr & p_0<0}\right\}
	\quad\matrix{\ll p|p'\rr> 0, \cr
	\one=\hbox{usual.}\cr}
&
\cr
\noalign{\hrule}
\superstrut
&
	G_F(x,y)
&&
	\omit\hfil $G_F = G_F\circ G_F$ \hfil
&&
	\omit\hfil$\matrix{N>0 & x^0 \lessgtr y^0\cr}$\hfil
&&
	{\rm see}\quad G^+\quad{\rm and}\quad G^-.
&
\cr
\noalign{\hrule}
\superstrut
&
	G^{(1)}(x,y)
&&
	\omit\hfil
	$\eqalign{&G= G^{(1)}\circ G^{(1)}\cr &G^{(1)}= G^{(1)}\circ G\cr
	&G^{(1)}=G^+\circ G^+-G^-\circ G^-\cr}$
	\hfil
&&
	\omit\hfil$\infty> N > -\infty$\hfil
&&
	p_0 \lessgtr 0 \quad \matrix{
	\ll p|p'\rr > 0,\cr
	\one=\hbox{unusual.}\cr}
&
\cr
\noalign{\hrule}
\superstrut
&
	G(x,y)
&&
	\omit\hfil $G= G\circ G$ \hfil
&&
	\omit\hfil?\hfil
&&
	p_0 \lessgtr 0\quad \matrix{
	\ll p|p'\rr {\rm indefinite},\cr
	\one=\hbox{usual.}\cr}
&
\cr
\noalign{\hrule}
\superstrut
&
	G_{NW}(x,y)
&&
	\omit\hfill $G_{NW}= G_{NW} \times G_{NW}$ \hfill
&&
	\eqalign{
	&\matrix{\hbox{(i) paths moving forward}\cr
	\hbox{in $x^0$, $N>0$}\cr}\cr
	&\matrix{\hbox{(ii) all paths not crossing}\cr
	\hbox{final surface, $N>0$}\cr}\cr}
&&
	p_0>0 \quad
	\matrix{\ll p|p'\rr> 0, \cr
	\one=\hbox{usual}\cr
	\hbox{non-relativistic.}\cr}
&
\cr
\noalign{\hrule\hrule}
}
}
}
$$
\par
\vfill
\eject
\hoffset=0in

%\voffset=8truein
\noindent{Figure 1. Paths for the non-relativistic propagator in the set
$p(\x',t'\to\x_t,t\to\x'',t'')$.}
%\eject
\vskip 1 cm
\noindent{Figure 2. Paths for the relativistic propagator.}
%\eject
\vskip 1 cm
\noindent{Figure 3. The surface $\Sigma$ divides the configuration space $\C$
into two components, $\C_1$ and $\C_2$. A path typically crosses $\Sigma$
many times; the point of first crossing is at $\u{x}_\s$.}
%\eject
\vskip 1 cm
\noindent{Figure 4. The path crosses the surface $\Sigma$ for the first
time at $\x=\x_\s$ and $t=t_\s$, and is in the set $p(\x',0\to \x_\s,t_\s
\to\x'',T)$.}
\vskip 1 cm
\noindent{Figure 5. A path crossing the surface $\Sigma$
and ending at $\x''$ is
cancelled by a path crossing the surface and ending at
$\x_{\s}+(\x_\s-\x'')$, provided that $V(\x)$ is independent of $x^0$.}
\vskip 3 cm
\noindent{FIGURES ARE AVAILABLE FROM THE AUTHORS}

\end